\documentclass[format=acmsmall, review=false, screen=true]{acmart}

\usepackage{booktabs} 
\usepackage{makecell}
\usepackage{lipsum}
\usepackage[utf8]{inputenc}
\usepackage{array}
\usepackage{wrapfig}
\usepackage{multirow}
\usepackage{tabularx}
\usepackage{pifont}
\usepackage{lineno}
\usepackage{balance}
\usepackage{xcolor,pifont}
\usepackage{longtable}
\usepackage{textcomp}

\usepackage{soul}

\usepackage{balance}
\usepackage[linesnumbered, ruled,vlined]{algorithm2e}
\RequirePackage{pdflscape}
\usepackage{subcaption}

\acmJournal{TOSEM}

\usepackage{flushend}
\usepackage{multicol}
\usepackage{placeins}
\usepackage{booktabs}
\usepackage{indentfirst}
\usepackage{fancybox}
\usepackage[most]{tcolorbox}
\usepackage{fontawesome}
\usepackage{mathtools}
\usepackage{MnSymbol,wasysym}
\usepackage{fontawesome}
\usepackage{rotating}
\usepackage{adjustbox}
\usepackage{colortbl}
\usepackage{color}
\usepackage{framed}
\definecolor{Gray}{gray}{0.9}
\definecolor{shadecolor}{gray}{0.95}
\usepackage{tikz}
\usetikzlibrary{trees,positioning,shapes,shadows,arrows}

\tikzset{
  basic/.style  = {draw, text width=2cm, drop shadow, font=\sffamily, rectangle},
  root/.style   = {basic, rounded corners=2pt, thin, align=center, fill=white},
  level-2/.style = {basic, rounded corners=6pt, thin,align=center, fill=white, text width=3cm},
  level-3/.style = {basic, thin, align=center, fill=white, text width=1.8cm}
}
\usepackage{tcolorbox}

\newcommand{\todo}[1]{}
\renewcommand{\todo}[1]{{\color{red} TODO: {#1}}}

\begin{document}

\title{Security Weaknesses of Copilot-Generated Code in GitHub Projects: An Empirical Study}


\author{Yujia Fu}
\affiliation{%
  \institution{School of Computer Science, Wuhan University}
  \country{China}
}
\email{yujia_fu@whu.edu.cn}

\author{Peng Liang}
\authornote{Corresponding author}
\affiliation{%
  \institution{School of Computer Science, Wuhan University}
  \country{China}
}
\email{liangp@whu.edu.cn}

\author{Amjed Tahir}
\affiliation{%
  \institution{Massey University}
  \country{New Zealand}
}
\email{a.tahir@massey.ac.nz}

 \author{Zengyang Li}
\affiliation{%
  \institution{School of Computer Science, Central China Normal University}
  \country{China}
}
\email{zengyangli@ccnu.edu.cn}

\author{Mojtaba Shahin}
\affiliation{%
  \institution{RMIT University}
  \country{Australia}
}
\email{mojtaba.shahin@rmit.edu.au}

\author{Jiaxin Yu}
\affiliation{%
  \institution{School of Computer Science, Wuhan University}
  \country{China}
}
\email{jiaxinyu@whu.edu.cn}

\author{Jinfu Chen}
\affiliation{%
  \institution{School of Computer Science, Wuhan University}
  \country{China}
}
\email{jinfuchen@whu.edu.cn}

\renewcommand{\shortauthors}{Fu et al.}

\begin{abstract}
Modern code generation tools utilizing AI models like Large Language Models (LLMs) have gained increased popularity due to their ability to produce functional code. However, their usage presents security challenges, often resulting in insecure code merging into the code base. Thus, evaluating the quality of generated code, especially its security, is crucial. \textcolor{black}{While prior research explored various aspects of code generation, the focus on security has been limited, mostly examining code produced in controlled environments rather than open source development scenarios. To address this gap, we conducted an empirical study, analyzing code snippets generated by GitHub Copilot and two other AI code generation tools (i.e., CodeWhisperer and Codeium) from GitHub projects. Our analysis identified 733 snippets, revealing a high likelihood of security weaknesses, with 29.5\% of Python and 24.2\% of JavaScript snippets affected. These issues span 43 Common Weakness Enumeration (CWE) categories, including significant ones like \textit{CWE-330: Use of Insufficiently Random Values}, \textit{CWE-94: Improper Control of Generation of Code}, and \textit{CWE-79: Cross-site Scripting}. Notably, eight of those CWEs are among the 2023 CWE Top-25, highlighting their severity. We further examined using \textit{Copilot Chat} to fix security issues in Copilot-generated code by providing \textit{Copilot Chat} with warning messages from the static analysis tools, and up to 55.5\% of the security issues can be fixed. We finally provide the suggestions for mitigating security issues in generated code.}
\end{abstract}



\maketitle



\begin{sloppypar}

\section{Introduction} \label{Introduction}
Code generation tools aim to automatically generate functional code based on prompts, which can include text descriptions (comments), code (such as function signatures, expressions, variable names, etc.), or a combination of text and code~\cite{sarkar2022like}. After writing an initial code or comment, developers can rely on code generation tools to complete the remaining code. This approach can save development time and accelerate the software development process. Automated code generation tools have always been a topic of active research~\cite{siddiq2022securityeval,li2022cctest}. Some of the earliest work can be traced back to the 1960s, when Waldinger and Lee proposed a program synthesizer called PROW, which automatically generated LISP programs based on specifications provided by users in the form of a predicate calculus~\cite{waldinger1969prow}. As computer languages continued to evolve, more and more programming languages began to support meta-programming, making automated code generation technology more efficient and flexible. In recent years, the rapid development of artificial intelligence technologies, particularly deep learning models, has accelerated the development of code generation technologies.

Recent advancements in code generation came with the emergence of Large Language Models (LLMs). LLMs are deep learning models trained on a large code/text corpus with powerful language understanding capabilities that can be used for tasks such as natural language generation, text classification, and question-answer systems \cite{chen2021evaluating}. Compared to previous deep learning methods, the latest developments in LLMs, such as Generative Pre-trained Transformer (GPT) models, have opened up new opportunities to address the limitations of existing automated code generation technology~\cite{mozannar2022reading}. Currently, LLM-based code generation tools have also been widely applied, such as Codex by OpenAI~\cite{codex}, AlphaCode by DeepMind~\cite{li2022competition}, and CodeWhisperer by Amazon~\cite{becker2022programming}. 

These models are trained on billions of public open-source lines of code, which include public code with unsafe coding patterns~\cite{he2023controlling}. Therefore, code generation tools based on such models can pose code quality issues~\cite{liu2023refining}, and the AI generated code may also suffer from security weaknesses~\cite{she2023pitfalls}. For example, GitHub Copilot may produce some insecure code, as its underlying Codex model is pre-trained on untrusted data from GitHub \cite{brown2020language}, which is known to contain buggy programs \cite{rokon2020sourcefinder}. According to the developer security company Snyk, GitHub Copilot may also replicate existing security weaknesses in code to suggest insecure code when the user's existing codebase contains security weaknesses~\cite{infoworld2024}.

In addition, the code with vulnerabilities generated by these code-generation tools may continue to be used to train the model, thus further generating code with vulnerabilities, leading to a vicious cycle.
Previous research has studied code generation tools, with more focus on the correctness of the results~\cite{lertbanjongngam2022empirical, wong2022exploring, dakhel2022github, pudari2023copilot}, and relatively less attention has been paid to security aspects~\cite{pearce2022asleep, siddiq2022empirical, perry2022users}. To the best of our knowledge, potential security weaknesses in practical scenarios have not been fully considered and addressed in previous work, and GitHub Copilot clarifies that ``\textit{the users of Copilot are responsible for ensuring the security and quality of their code}''~\cite{githubcopilot}. Code generation algorithms of Copilot are incentivized to suggest code to be accepted rather than other code qualities, e.g., easy to read and understand, which has an adverse impact on the code quality generated by Copilot \cite{harding2024coding}. GitHub also provides tools such as CodeQL to help developers scan for security weaknesses in their code.

\textcolor{black}{To this end, we conducted an empirical study on the security weaknesses of the generated code by GitHub Copilot and other AI code generation tools (CodeWhisperer and Codeium), which is available on GitHub. We chose Copilot as our main research subject because it is a commercial instance of AI programming and has gained much attention and popularity among developers since its launch in 2021. The security weaknesses of code generated by Copilot have also gained attention in the research and practice community. Furthermore, thousands of developers in the GitHub community have shared their experiences of using Copilot in open source projects~\cite{dakhel2022github}. 
We also include the code generated by other two popular AI code generation tools, CodeWhisperer and Codeium, for the purpose of increasing the generalizability of this research.}
We then collected the code generated by Copilot and other generation tools that has been used in projects on GitHub and analyzed the security of the generated code through the lens of open source development environment. Then, we used static analysis tools to perform security analysis on the collected code snippets and classified the security weaknesses in the code snippets using the Common Weakness Enumeration (CWE).

\textcolor{black}{\textbf{Our findings show that}: (1) around 30\% of code snippets have security weaknesses; (2) the security weaknesses are diverse and related to 43 different CWEs, in which\textit{CWE-330: Use of Insufficiently Random Values}, \textit{CWE-94: Improper Control of Generation of Code ('Code Injection')}
and \textit{CWE-79: Cross-site Scripting} and are the most frequently occurred; and (3) among the 38 CWEs identified, eight CWEs belong to the currently recognized 2023 CWE Top-25. Six CWEs belong to Stubborn Weaknesses in the CWE Top 25. (4) It is possible to use \textit{Copilot Chat} to fix security issues in Copilot-generated code, but the success of \textit{Copilot Chat}'s fixes varies between CWEs. (5) Providing \textit{Copilot Chat} with a warning message from the static analysis tool resulted in a better fix. Using the \/fix command can fix 19.3\% of security issues, while using the enhanced prompt raises the rate to 55.5\%.}

\textbf{Contributions}: (1) We curated a dataset of code snippets generated by Copilot that has been used in projects on GitHub (a curated data~\cite{dataset} is made available online ) and conducted security checks on them, which can to some extent reflect the frequency of security weaknesses encountered by developers when using Copilot to generate code in actual coding. In addition to this, we also categorized the application areas of these code snippets. (2) We extensively checked all possible CWEs in the code snippets and analyzed them. This can help developers understand the common CWEs caused by using Copilot to generate code in actual coding and how to safely accept the code suggestions provided by Copilot. (3) We utilized \textit{Copilot Chat} to fix code snippets generated by Copilot, demonstrating Copilot's ability to fix security issues to a certain extent.

The rest of this paper is structured as follows: Section \ref{Related Work} presents the background and related work. Section \ref{Research Design} presents the research questions and the research design of this study. Section \ref{Results} presents the results of our study, which are further discussed in Section \ref{Discussion}. The potential threats to validity are clarified in Section \ref{Threats to Validity}. Section \ref{Conclusions and Future Work} concludes this work with future work directions.

\section{Background and Related Work} \label{Related Work}
In this section, we present the related work in four aspects, i.e., AI code generation tools (Section~\ref{AIAssistedCodeTools}), security of code generation techniques and LLMs (Section~\ref{SecurityofCodeGenerationTechniquesandLLMs}), \textcolor{black}{how AI code generation tools work (Section~\ref{How AI code generation works})}, and static analysis tools for analyzing security weaknesses (Section~\ref{StaticAnalysisToolsforSecurityScanning}).
    
\subsection{AI Code Generation Tools}\label{AI Code Generation Tools} \label{AIAssistedCodeTools}
With the rise of code generation tools integrated with IDEs, many studies have evaluated these systems based on transformer models to better understand their effectiveness in open source development scenarios. Previous research mainly focused on whether the code generated by these tools can meet users' functional requirements. 
Yetistiren \textit{et al.}~\cite{yetistiren2022assessing} evaluated the effectiveness, correctness, and efficiency of the code generated by GitHub Copilot, and the results showed that GitHub Copilot could generate valid code with a success rate of 91.5\%, making it a promising tool. Sobania \textit{et al.}~\cite{sobania2022choose} evaluated the correctness of the code generated by GitHub Copilot and compared the tool with an automatic program generator with a Genetic Programming (GP) architecture. They concluded there was no significant difference between the two methods on benchmark problems. Nguyen and Nadi~\cite{nguyen2022empirical} conducted an empirical study using 33 LeetCode problems and created queries for Copilot in four different programming languages. They evaluated the correctness and comprehensibility of the code suggested by Copilot by running tests provided by LeetCode. They found that Copilot's suggestions have lower complexity. Burak \textit{et al.}~\cite{yeticstiren2023evaluating} evaluated the code quality of AI code generation tools. They compared the improvements between the latest and older versions of Copilot and CodeWhisperer and found that the quality of generated code had improved.

In recent years, researchers have also started to focus on the experience of developers when using AI code generation tools and how the tools can improve productivity by observing their behavior. 
Vaithilingam \textit{et al.}~\cite{vaithilingam2022expectation} studied how developer use and perceive Copilot and found that although Copilot may not necessarily improve task completion time or success rate, it often provides a useful starting point. They also noted that the participants had difficulties understanding, editing, and debugging the code snippets generated by Copilot. Barke \textit{et al.}~\cite{barke2022grounded} presented the first theoretical analysis of how developer interact with Copilot based on the observations of 20 participants. Sila \textit{et al.}~\cite{lertbanjongngam2022empirical} conducted an empirical study on AlphaCode, identifying similarities and performance differences between code generated by code generation tools and code written by human developers. They argued that software developers should check the generated code for potentially problematic code that could introduce performance weaknesses.

The studies presented above have conducted a relatively extensive evaluation of code-generation tools regarding correctness, effectiveness, and robustness. However, its security still has room for improvement, as detailed below. 

\subsection{Security of Code Generation Techniques and LLMs}\label{SecurityofCodeGenerationTechniquesandLLMs}
Code security is an issue that cannot be ignored in the software development process. Recent work has primarily focused on evaluating the security of the code generation tools and the security of the LLMs based on these tools. 

Pearce \textit{et al.}~\cite{pearce2022asleep} first evaluated the security of GitHub Copilot in generating programs by identifying known weaknesses in the suggested code. The authors prompted Copilot to generate code for 89 cybersecurity scenarios and evaluated the weaknesses in the generated code. They found that 40\% of the suggestions in the relevant context contained security-related bugs (i.e., CWE classification from MITRE~\cite{cwe}). Siddiq \textit{et al.}~\cite{siddiq2022empirical} conducted a large-scale empirical study on code smells in the training set of a transformer-based Python code generation model and investigated the impact of these harmful patterns on the generated code. They observed that Copilot introduces 18 code smells, including non-standard coding patterns and two security smells (i.e., code patterns that often lead to security defects). Khoury \textit{et al.}~\cite{khoury2023secure} studied the security of the source code generated by the ChatGPT chatbot based on LLMs and found that ChatGPT was aware of potential weaknesses but still frequently generated some non-robust code. Elgedawy \textit{et al.}~\cite{elgedawy2024ocassionally} compared the capabilities of four code generation models using nine code generation tasks. They collected 61 code outputs and studied their security. The results revealed that the code generated by different LLMs exhibited disparate levels of security robustness.

Several researchers also compared the situation in which code generation tools produce insecure code with that of human developers. Sandoval \textit{et al.}~\cite{sandoval2022security} conducted a security-driven user study, and their results showed that the rate at which AI user programming produced critical security errors was no more than 10\% of the control group, indicating that the use of LLMs does not introduce new security risks. Asare \textit{et al.}~\cite{asare_is_2023} conducted a comparative empirical analysis of these tools and language models from a security perspective and investigated whether Copilot is as bad as humans in generating insecure code. They found that while Copilot performs differently across vulnerability types, it is not as bad as human developers when introducing code vulnerabilities.
 In addition, researchers have also constructed datasets to test the security of these tools. Tony \textit{et al.}~\cite{tony2023llmseceval} proposed LLMSecEval, a dataset containing 150 natural language prompts that can be used to evaluate the security performance of LLMs. Siddiq \textit{et al.}~\cite{siddiq2022securityeval} provided a dataset, SecurityEval, for testing whether a code generation model has weaknesses. The dataset contains 130 Python code samples. Natella \textit{et al.}~\cite{natella2024ai} provide a dataset and conduct an experimental evaluation of the security of three popular LLMs (GitHub Copilot, Amazon CodeWhisperer, and CodeBERT).

\textcolor{black}{Different from prior work, we studied the security weaknesses exhibited by code generation tools in open source development environment (i.e., GitHub). We collected code snippets from GitHub generated by developers using Copilot in daily production as a source of research data, whereas in the study by Pearce \textit{et al.}~\cite{pearce2022asleep}, the research data came from code generated by the authors using Copilot based on natural language prompts related to high-risk network security weaknesses.} Additionally, Pearce \textit{et al.} configured CodeQL only to examine CWEs targeted by security weaknesses associated with the provided scenarios. In contrast, we used various static analysis tools to examine all types of CWEs and analyze them extensively. Our research results may help developers understand what common CWEs are prone to result from using Copilot to generate code in real coding.

\begin{figure}
    \captionsetup{labelfont={color=black}, textfont={color=black}}
    \begin{subfigure}{0.65\textwidth}
        \includegraphics[width=\linewidth]{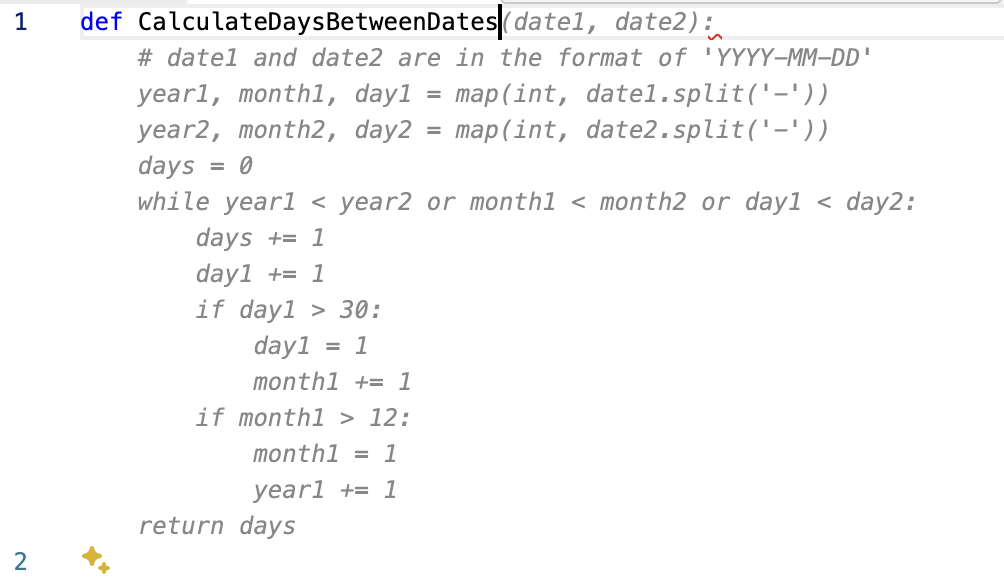}
        \caption{Method 1: Copilot completes a function using the function name}
        \label{fig:function}
    \end{subfigure}

    \begin{subfigure}{0.89\textwidth}
        \includegraphics[width=\linewidth]{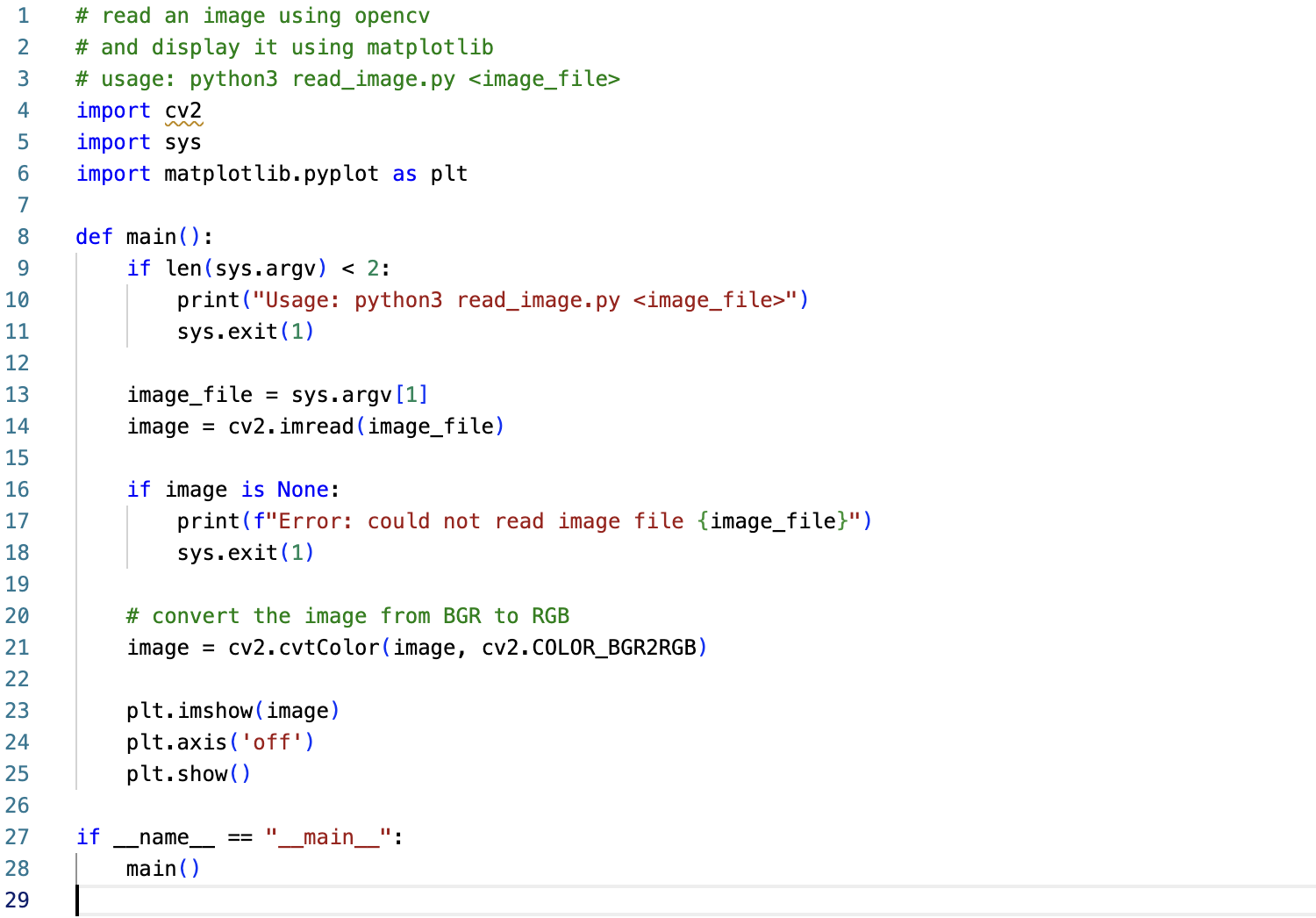}
        \caption{Method 2: Copilot implements the function in the comment}
        \label{fig:comment}
    \end{subfigure}
    
    \caption{Two methods of using Copilot in action}
    \label{fig:copilot-example}  
    \captionsetup{labelfont={}, textfont={}}  
\end{figure}

\begin{figure*}[htbp]
        \includegraphics[width=0.99\linewidth]{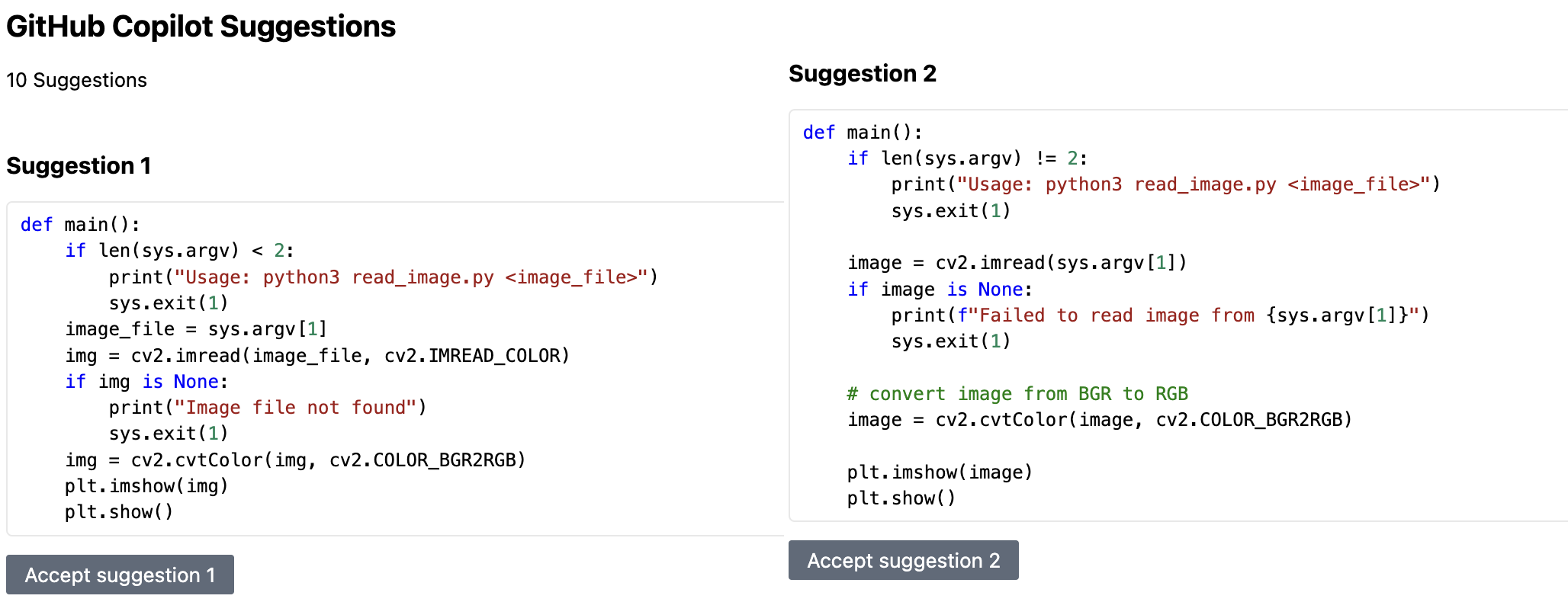}
        \captionsetup{labelfont={color=black}, textfont={color=black}}
        \caption{More suggestions provided by Copilot}
        \captionsetup{labelfont={}, textfont={}}
        \label{fig:suggestions}
\end{figure*}

\textcolor{black}{
\subsection{How AI Code Generation Tools Work: Copilot as an Example}\label{How AI code generation works}
Existing automated code generation tools, such as Copilot, CodeWhisperer, and Codeium, are often integrated into IDEs. These tools provide developer with real-time code suggestions based on given prompts. Those suggestions can range from code snippets to complete functions. We use GitHub Copilot as an example to demonstrate how these tools are used.}
\textcolor{black}{When developers write code in their IDE, Copilot continuously scans the program and performs the following actions~\cite{copilot-usage}: (1) complete existing code based on function or variable names; (2) generate code suggestions from inline natural language comments; (3) automatically fill in repetitive code based on the context of existing code; and (4) generate test cases. Figure~\ref{fig:copilot-example} shows an example of Copilot usage in action. In Figure~\ref{fig:function}, Copilot automatically suggests the entire function body (shown in gray text) when the function name is provided. A user can then accept the suggestion if desired. In Figure~\ref{fig:comment}, a function description is given in the comments (Lines 1-2), and Copilot provides code suggestions starting from Line 3. After accepting a suggestion, Copilot will continue to suggest the next line of code. A user can also view alternative suggestions - Copilot will provide 10 different suggestions at a time, as shown in Figure~\ref{fig:suggestions}.
}

\subsection{Security Static Analysis}\label{StaticAnalysisToolsforSecurityScanning}
Detecting software vulnerabilities is critical to improve security and ensure quality ~\cite{iannone2022secret}. Vulnerabilities can be detected through static, dynamic, or hybrid analysis. Static analysis offers greater coverage and allows users to analyze programs without the need to execute them, but it can be unsound and imprecise ~\cite{sui2020recall}. On the other hand, dynamic analysis is more sound and precise (as it captures actual program behaviour at runtime)  but can be incomplete and lacks coverage~\cite{ernst2003static}. Due to their ease of use and low cost compared to dynamic analysis, static analysis tools are widely used for security analysis \cite{nunes2018comment,huth2019static}. 

There are several security static analysis tools available for different programming languages and application types. OWASP and Snyk~\cite{owasp, snyk_code} provide a list of commonly used static analysis tools for security analysis. This includes tools like CodeQL, a general-purpose automatic analysis tool; FindBugs/SpotBugs for Java; ESLint for JavaScript programs; Bandit for Python; and GoSec for Go programs. Those tools have been widely used in previous security analysis research ~\cite{siddiq2022empirical, tomasdottir2018adoption, pearce2022asleep}. 

Kaur \textit{et al.}~\cite{kaur2020comparative} compared static analysis tools for vulnerability detection in analyzing C/C++ and Java source code. Tomasdottir \textit{et al.}~\cite{tomasdottir2018adoption} conducted an empirical study on ESLint, the most commonly used JavaScript static analysis tool among developers. Pearce \textit{et al.}~\cite{pearce2022asleep} used CodeQL to scan security weaknesses in the generated Python and C++ code. Siddiq \textit{et al.}~\cite{siddiq2022securityeval} used Bandit to check Python code generated using a test dataset. Lisa \textit{et al.}~\cite{do2020software} reported on users' goals, motivations, and strategies when using static analysis tools.

These static analysis tools support different analysis algorithms and techniques. By using multiple tools for analysis, potential weaknesses in the code can be discovered from different perspectives and levels, avoiding omissions and improving the accuracy of the analysis. Our study first used CodeQL to scan the collected code snippets. CodeQL is an open-source tool that supports multiple languages, including Java, JavaScript, C++, C\#, and Python. It can find weaknesses in a codebase based on known weaknesses/rules. In addition, to obtain more comprehensive analysis results, we supplemented the scan of code in different languages with static analysis tools (i.e., Cppcheck and Bandi) tailored to specific languages.

\section{Research Design} \label{Research Design}
In this section, we describe our research design in detail. In Section~\ref{Goal and Research Questions}, we first define our Research Questions (RQs), followed by the process of collecting and filtering the code snippets generated by code generation tools such as Copilot in Section~\ref{Data Collection and Filtering}. We then explain the security analysis performed on the identified snippets and the process of filtering the raw results generated by static analysis tools in Section~\ref{Data Analysis and Results Filtering}. Finally, in Section~\ref{security weakness Mitigation}, we introduce the process of using \textit{Copilot Chat} to fix the security weaknesses identified by the static analysis tools.

\subsection{Research Goal and Questions}\label{Goal and Research Questions}
\textcolor{black}{This study aims to understand the potential security weaknesses in AI-generated code produced in open source GitHub projects. We first collect code snippets generated by code generation tools such as Copilot from GitHub projects, which is our data source. It should be noted that it is not possible to access all the AI-generated code in GitHub projects, as there is no direct way to identify if part of a file was generated by code generation tools such as Copilot (i.e., source files do not contain any signatures to indicate if Copilot generates the code). However, we can identify many code snippets by searching the repository description and the comments provided in the code (see the details in Section \ref{Filtering Code Snippets}).}

We first used keyword-based search to collect projects and files containing code generated by Copilot and other AI code generation tools (CodeWhisperer and Codeium) from GitHub and filtered them manually. We subsequently analyzed the functionality and application domains of the collected code snippets to get a comprehensive understanding of the dataset. Next, we performed a static security analysis on the identified code. 

After running the analysis and obtaining the analysis results, we manually checked the results to remove false positives reported by the static analysis tools. We then used CWEs to classify the filtered results for further analysis to answer our Research Questions (RQs). \textcolor{black}{In addition, we used \textit{Copilot Chat}~\cite{copilot_chat} to fix code snippets containing security weaknesses. \textit{Copilot Chat} is an interactive tool launched by OpenAI based on the GPT-4 model, designed to enable natural language-driven coding as part of Copilot. Developers can use \textit{Copilot Chat} to perform tasks such as code analysis and fixing bugs. We used the warnings from static analysis tools as prompts and fed them to \textit{Copilot Chat} in order to get the fixes. We then performed a security analysis on the fixed code to evaluate the effectiveness of those fixes. }

\begin{figure*}[htbp]
	\centering
	\includegraphics[width=\linewidth]{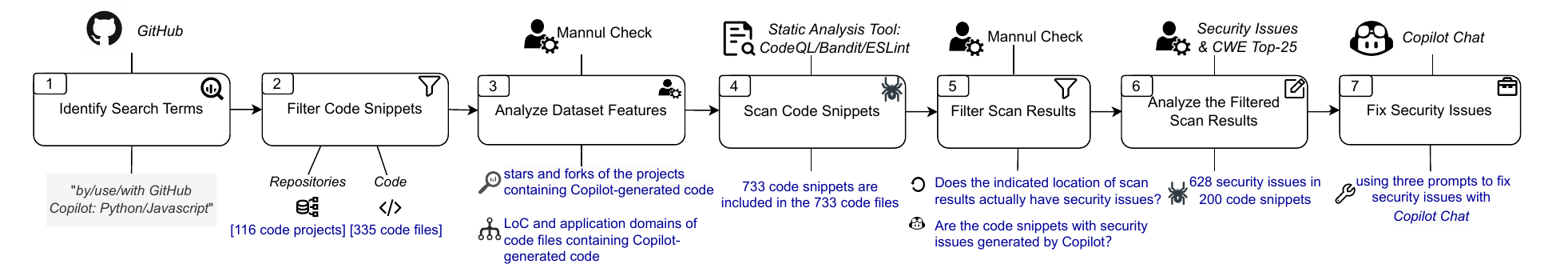}
        \captionsetup{labelfont={color=black}, textfont={color=black}}
	\caption{Overview of the research process}
    \captionsetup{labelfont={}, textfont={}}
	\label{fig:Overview of research process}
\end{figure*}

To conduct our empirical study, we followed the guidelines of Easterbrook \textit{et al.}~\cite{easterbrook2008selecting}. The RQs, their rationale, and the research process of this study (see Figure~\ref{fig:Overview of research process}) are detailed in the following subsections.

\noindent\textbf{RQ1. How secure is the code generated by Copilot in GitHub projects? }

\noindent\textbf{Rationale:} Copilot may produce code suggestions that developers accept, but these suggestions may include security weaknesses that could make the program vulnerable. The answer to RQ1 helps understand the frequency of security weaknesses developers encounter when using Copilot in production. We also include the code generated by other two popular AI code generation tools, CodeWhisperer and Codeium, in order to increase the generalizability of this research.\\

\noindent\textcolor{black}{\textbf{RQ2. What security weaknesses are present in the code generated by Copilot across different application domains?}}

\noindent\textcolor{black}{\textbf{Rationale:} Copilot-generated code may contain security weaknesses~\cite{pearce2022asleep, majdinasab2024assessing}, and developers should conduct a rigorous security review before accepting the generated code suggestions. Copilot's documentation notes the following: ``\textit{users of Copilot are responsible for ensuring the security and quality of their code}''~\cite{githubcopilot}. By identifying common or recurring weaknesses, developers can be more prepared to prevent, mitigate, or fix these security weaknesses. The answer to RQ2 can help developers better understand possible security weaknesses in the code generated by Copilot and other AI code generation tools, as well as the characteristics of code that contains certain types of security weaknesses, allowing them to be vigilant about the generated code before it is integrated into their code base.}\\



\noindent\textcolor{black}{\textbf{RQ3. Can \textit{Copilot Chat} help fix the security weaknesses found in the code generated by Copilot?}}

\noindent\textcolor{black}{\textbf{Rationale:} The purpose of \textit{Copilot Chat} is to assist developers by answering questions about their code or providing fixes for issues identified in the code. \textit{Copilot Chat} is integrated with Copilot to form a powerful AI assistant that can help developers build at their speed of thought in the natural language of their choice~\cite{github_copilot_chat}. It is claimed that \textit{Copilot Chat} can assist in fixing security weaknesses. 
In this RQ, we utilize \textit{Copilot Chat} to examine whether it can help fix the security weaknesses identified in the Copilot-generated code because most of the security weaknesses in RQ1 and RQ2 come from Copilot. Our approach is to utilize \textit{Copilot Chat} in the loop by \textit{collect generated code \textrightarrow analyze the code for security weaknesses \textrightarrow check for fixes in \textit{Copilot Chat} \textrightarrow re-analyze the code to check if the fix is correct}. Answering RQ3 can help to understand the capability of \textit{Copilot Chat} in fixing Copilot-generated code that contains security issues.}

\subsection{Data Collection and Filtering}\label{Data Collection and Filtering}
We chose GitHub as the primary data source for answering our RQs. As the world's largest code hosting platform, GitHub contains millions of public code repositories and offers access to a large number of code resources, allowing us to cover multiple programming languages and project types in our study~\cite{cosentino2017systematic}. At the same time, thousands of developers in the GitHub community have shared their experiences of using Copilot in open-source projects. We identified code snippets generated by AI code generation tools from GitHub projects.

\textcolor{black}{We chose Copilot as our main research subject as it is a well-known commercial instance of an AI pair programming tool that has gained increased popularity among developers since its launch in 2021. Copilot is one of the most popular general-purpose AI code generation tools and only second to ChatGPT in terms of overall AI assistance tools developers use in practice~\cite{stackoverflow2024survey}.} 
\textcolor{black}{The security drawbacks of code generated by Copilot have been investigated in previous studies \cite{pearce2022asleep, gitguardian2021github, mirai_copilot_security}. 
To increase the generalizability of the research, we also attempted to collect data from two popular AI code generation tools, CodeWhisperer and Codeium - both are general-purpose, AI-driven code generators. We did not collect data from specialized AI code generation tools, such as AlphaCode as it focuses on solving complex coding tasks rather than everyday code generation.}

\subsubsection{Code Snippets Collection} 
\textbf{Step 1.} \textcolor{black}{We utilized GitHub REST API to automatically collect projects and files containing AI-generated code.} We started with a pilot search on GitHub to formulate our search keywords. First, we used the names of the code generation tools, such as ``GitHub Copilot'' and ``CodeWhisperer'' as the basic search terms. 
It should be noted that many code snippets in the search results contained the ``GitHub Copilot'' keyword, but the keyword was not used to declare that the code snippets were generated by Copilot. Developers may use them to describe their experience of using Copilot to generate code or showcase information related to Copilot. These code snippets are not what we seek as they do not directly relate to the code generated by Copilot. \textcolor{black}{By manually analyzing the search results, we found that restricting the search terms provided more accurate results. We decided to use keyword combinations such as \{\textit{by, use, with}\} + \{\textit{GitHub Copilot, CodeWhisperer, Codeium}\} as search keywords. This approach enabled us to focus more on the code generated using AI code generation tools (e.g., Copilot) rather than code snippets that contain other content related to the generation tools.}

\textcolor{black}{In addition, we further limited the types of code snippets during the search to Python and JavaScript. Both are dynamically typed languages where a full analysis can be conducted without the need for compilation. On the other hand, statically typed languages such as Java and C++ require the code to be compiled first to generate the database required for analysis. Our data collection method makes it difficult to properly compile individual code snippets, given that some code is unavailable when conducting the analysis.}

\textcolor{black}{We selected Python and JavaScript as they are currently the two most popular programming languages used by developers~\cite{octoverse2024}, and the most frequently used with GitHub Copilot~\cite{zhang2023demystifying}. The popularity of the two languages ensures that we can collect a substantial amount of generated code from GitHub projects, providing a significantly larger sample size and statistically meaningful results compared to other languages.
We collected the search results under the filter \textit{Code} with file names, file paths, and forks and stars of the repositories where the files were located. Considering that some projects declared using GitHub Copilot generated code in their README files or project descriptions provided in GitHub, we decided to retain the results from the filter \textit{Repo} with project name, project path, project description, and forks and stars of the projects.}




\textcolor{black}{Table \ref{Search terms on Github} shows a breakdown of the search terms we used and the number of search results per language. The same search result may contain multiple keywords, meaning there are duplicate projects in the collected data. After removing duplications, we ended up with 3,589 search results, of which 3,141 were from the \textit{Code} label and 445 from the \textit{Repository} label. }

\begin{table}[h]
\footnotesize
\captionsetup{labelfont={color=black}, textfont={color=black}}
\caption{Search results based on different terms}
\captionsetup{labelfont={}, textfont={}}  
\label{Search terms on Github}
\centering
\begin{tabularx}{\textwidth}{ m{2.5cm}<{\centering} m{1.8cm}<{\centering} m{1.8cm}<{\centering} m{1.8cm}<{\centering} m{1.8cm}<{\centering} m{2.3cm}<{\centering} }
\toprule
\textbf{Search Term} & \textbf{Code (Py)} & \textbf{Code (Js)} & \textbf{Repo (Py)} & \textbf{Repo (Js)} & \textbf{Total} \\
\midrule
by GitHub Copilot     & 1320 & 1256 & 48  & 43  & \textbf{2667} \\
use GitHub Copilot    & 1344  & 1384  & 84  & 93  & \textbf{2905} \\
with GitHub Copilot   & 1532 & 1876 & 95  & 177 & \textbf{3680} \\
\midrule
by CodeWhisperer     & 39   & 44   & 0   & 1   & \textbf{84}   \\
use CodeWhisperer    & 41   & 46   & 4   & 2   & \textbf{93}   \\
with CodeWhisperer   & 68   & 39   & 4   & 0   & \textbf{111}  \\
\midrule
by Codeium           & 85   & 73   & 0   & 1   & \textbf{159} \\
use Codeium          & 31   & 26   & 0   & 1   & \textbf{58}   \\
with Codeium         & 99   & 69   & 0   & 2   & \textbf{170}  \\
\midrule
\textbf{Total}       & \textbf{4559} & \textbf{4813} & \textbf{235} & \textbf{320} & \textbf{9927} \\
\bottomrule
\end{tabularx}

\end{table}

\subsubsection{Filtering Code Snippets}\label{Filtering Code Snippets}
\textbf{Step 2.} After obtaining the data from the keyword searches, we further filtered them by not only considering the accuracy of the keywords but also by investigating the project's documentation, code comments, and other metadata in the search results to determine whether they were generated by GitHub Copilot and other code generation tools. \textcolor{black}{Additionally, since we aimed to obtain code snippets used in open-source GitHub projects, we manually excluded search results related to solving simple programming practice problems on platforms that provide foundational exercises for enhancing algorithmic thinking and coding skills, such as bisection lookup and fast sorting tasks from LeetCode. We consider these programming exercises to be more closely related to correctness than security.}

\begin{figure*}[htbp]
	\centering
	\includegraphics[width=0.9\textwidth]{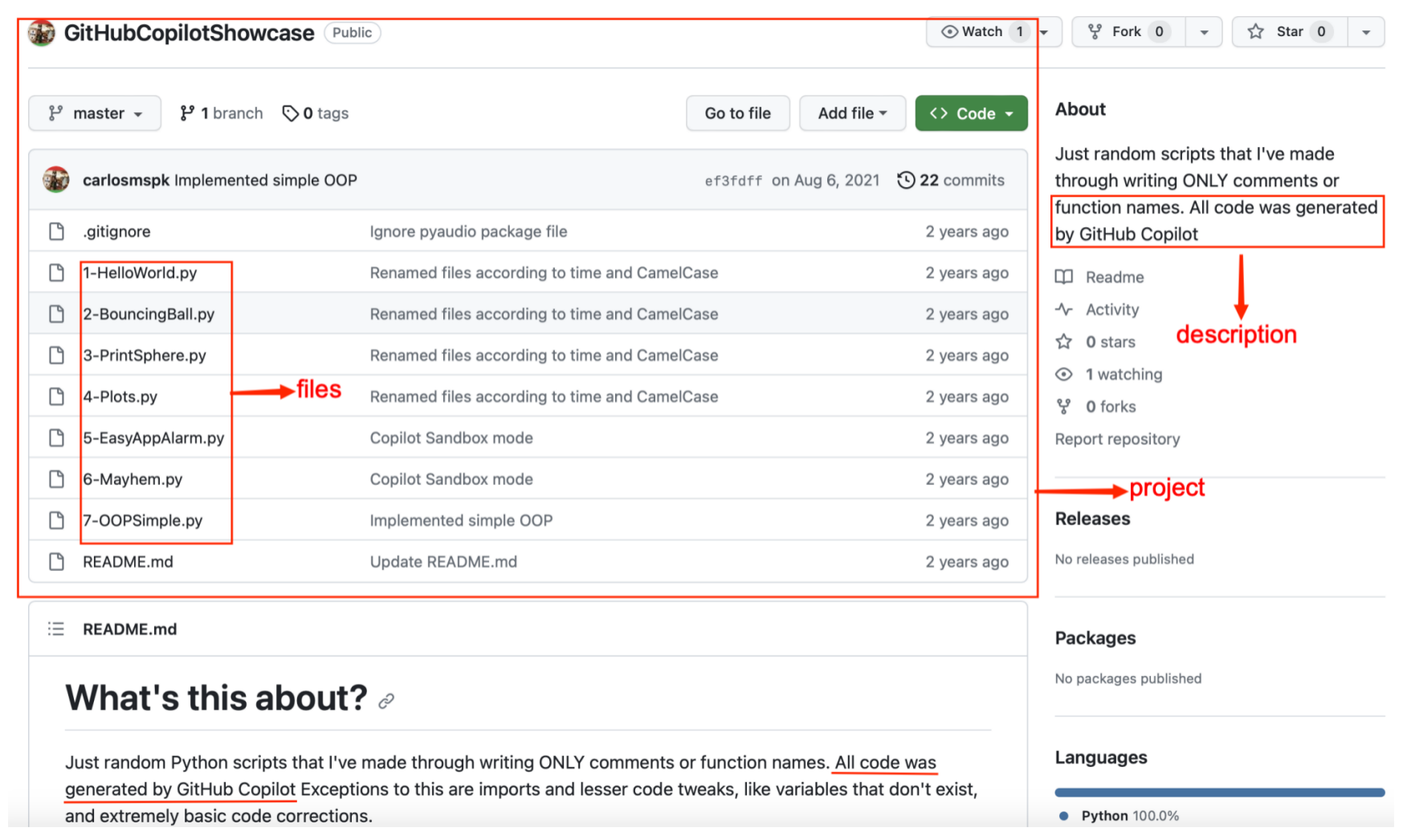}
	\caption{Example of \textit{Inclusion Criterion 1}: projects fully written by Copilot}
	\label{fig:Example of rule 1}
\end{figure*}

\begin{figure*}[htbp]
	\centering
	\includegraphics[width=0.7\textwidth]{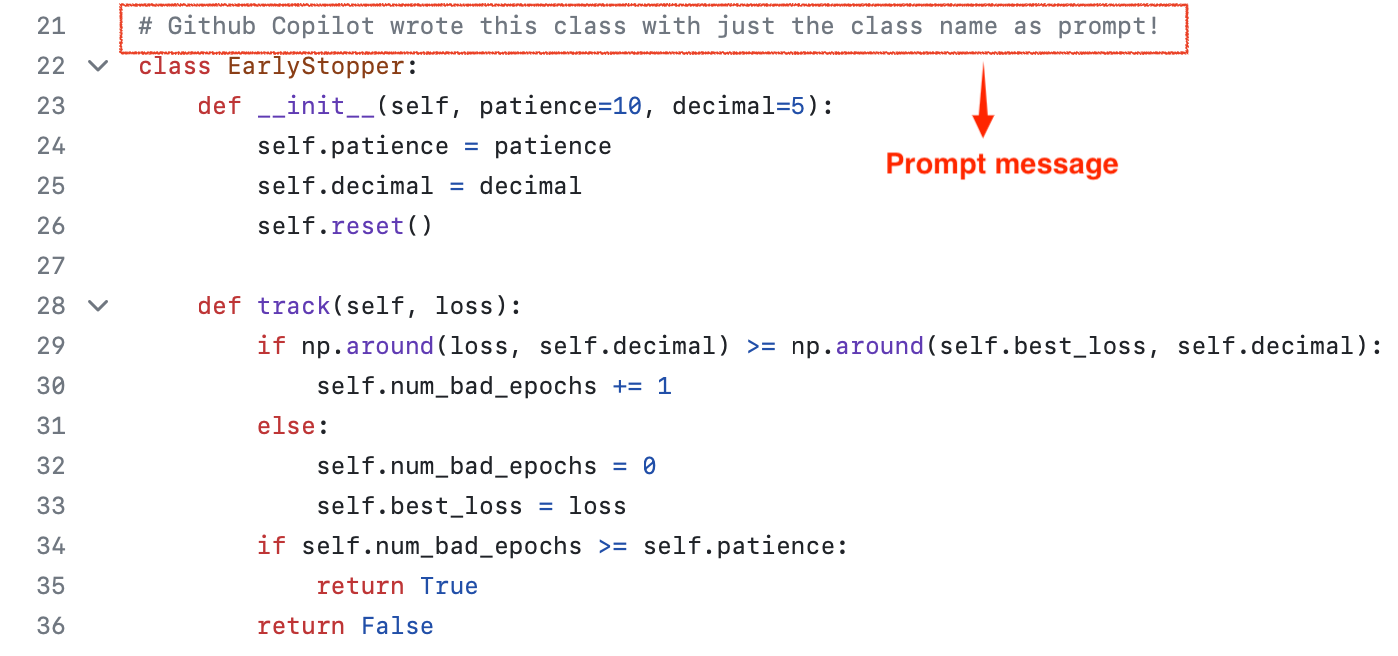}
	\caption{Example of \textit{Inclusion Criterion 2}: files with comments showing the code generated by Copilot}
	\label{fig:Example of rule 2}
\end{figure*}

\begin{figure*}[htbp]
	\centering
	\includegraphics[width=0.7\textwidth]{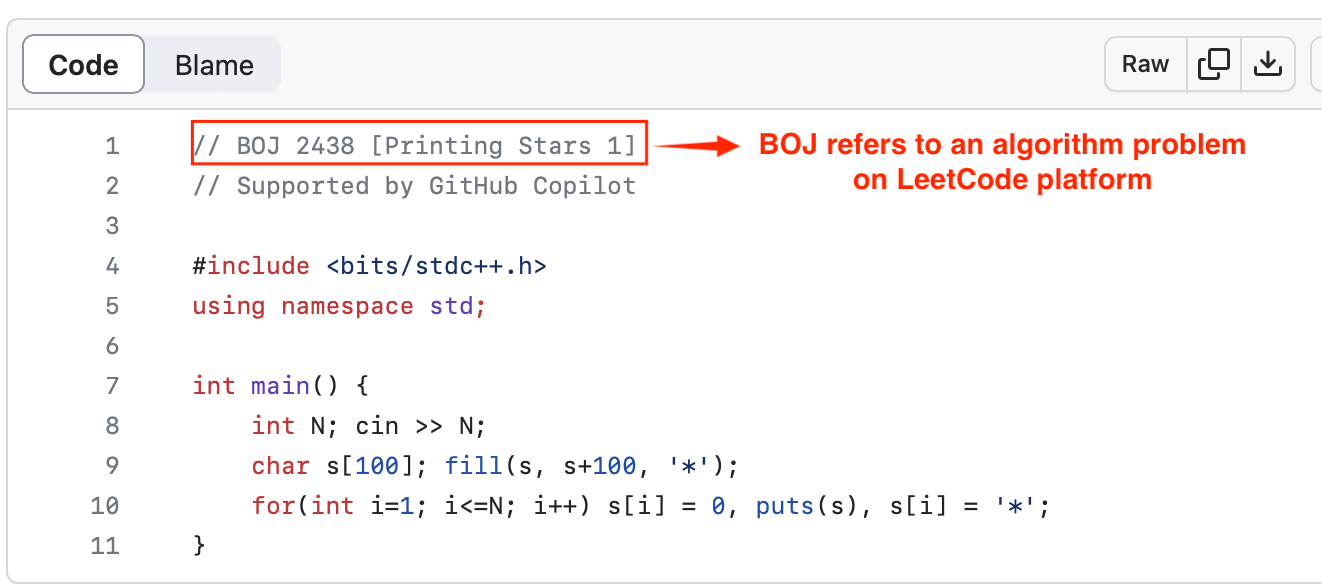}
	\caption{\textcolor{black}{Example of \textit{Exclusion Criterion 1}: files used to solve simple algorithm problems}}
	\label{fig:Example of rule 3}
\end{figure*}

We begin by explaining the terminology used in data filtering: the search results under the \textit{Repository} label are the projects that contain code files, and the search results under the \textit{Code} label are individual code files. Those code files contain code snippets generated by Copilot. In filtering the projects, we followed \textbf{three criteria} including two \textit{inclusion criteria} and one \textit{exclusion criterion}. \textit{Inclusion Criterion 1}: For search results under the \textit{Repository} label, we identified projects that are fully generated by Copilot, as declared in the project description or the associated README file(s). We retained code files for Python and JavaScript. \textit{Inclusion Criterion 2}: For search results under the \textit{Code} label, we retained code files with comments showing the code generated by Copilot. \textcolor{black}{\textit{Exclusion Criterion 1}: We excluded code files used to solve simple programming practice problems. Simple programming practice problems mainly refer to those standard coding exercises on platforms such as LeetCode, which are often named BOJ-number.} We provide examples for the three criteria in Figure~\ref{fig:Example of rule 1}, Figure~\ref{fig:Example of rule 2}, and Figure~\ref{fig:Example of rule 3}. 
As shown in the example in Figure~\ref{fig:Example of rule 1} for the \textit{Repository} label, we kept all the Python files. In the next example in Figure~\ref{fig:Example of rule 2}, we kept the entire file where the Copilot-generated code snippet was located. In Figure~\ref{fig:Example of rule 3}, the code snippet was removed as it was determined that the code just solved a simple algorithmic problem. 

\textcolor{black}{We manually filtered all the data, labeling whether they were generated by Copilot and other code generation tools based on the three criteria mentioned above. Two authors independently completed this manual filtration process for all the data within a period of two weeks. The first author compared the labeling results by the two coders and calculated the level of agreement between them using Cohen’s Kappa coefficient~\cite{Jacob1960Coefficient}. Cohen’s Kappa coefficient was 0.84, which was higher than 0.8, indicating a high level of agreement between the two coders and ensuring a good accuracy of the labeling results. Besides, for those results where the two authors disagreed on the labeling, they resolved the disagreements through a negotiated agreement approach~\cite{campbell2013coding} during a meeting. After manual filtration, we retained 116 projects under the \textit{Repository} label and 335 code files under the \textit{Code} label.}


For the code files retained under the \textit{Repository} label, we consider the entire code file as code generated by Copilot and other code generation tools. In other words, we assume that Copilot and other code generation tools generate all the code in the file because it was stated in the README file that it was all generated by Copilot and other code generation tools. For code files retained under the \textit{Code} label, we know that the files contain code snippets, perhaps even just a few lines of code, generated by Copilot and other code generation tools. Instead of identifying the specific AI-generated code in this step, we combine the warning messages from the security analysis and the code comments in the file to determine whether Copilot and other code generation tools generate the code snippet with the security problem (this process is explained further in Section \ref{Results Filtering}). As a result, our dataset consisted of 733 distinct code files, with 672 from Copilot, 38 from CodeWhisperer, and 23 from Codeium. Table \ref{Code Snippets from Code} gives the type and number of code files. A curated dataset, comprising all the data collected during the research process, has been made available~\cite{dataset}.

\begin{table}[ht]
\footnotesize
\caption{Code snippets from GitHub}
\label{Code Snippets from Code}{

\begin{tabular}{llccc}
\toprule
\textbf{\#} & \textbf{Language} & \textbf{\# Code Snippets: Repository} & \textbf{\# Code Snippets: Code} & \textbf{Total}  \\ \midrule
\textbf{L1}          & Python        & 190  & 229   & 419 \\
\textbf{L2}          & JavaScript    & 208  & 106   & 314 \\ \hline
\textbf{Total}       &               & 398  & 335   & \textbf{733} \\ \bottomrule
\end{tabular}
}
\end{table}

\begin{figure}[htbp]
    \centering
    \includegraphics[width=0.99\textwidth]{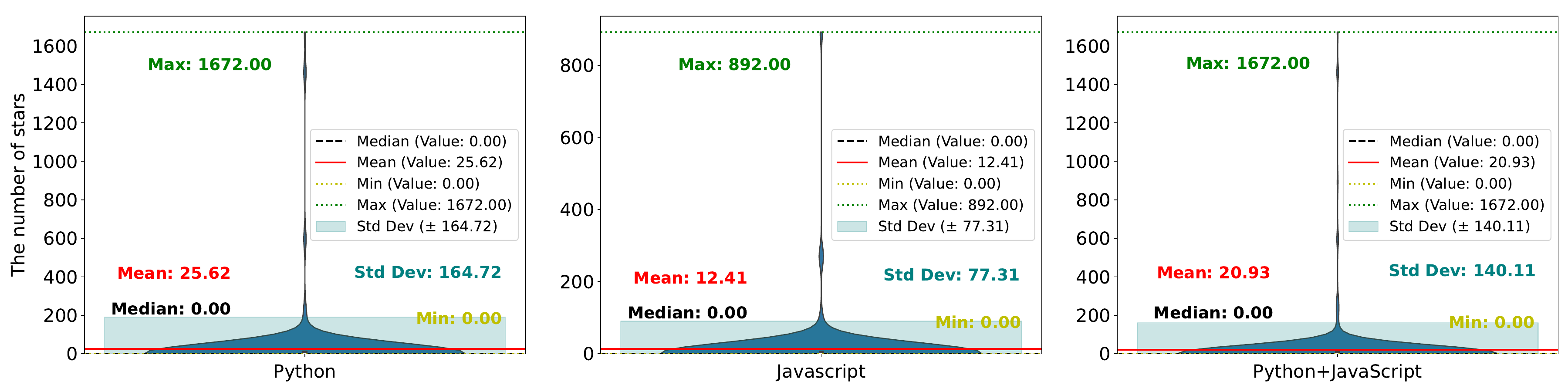}
    \captionsetup{labelfont={color=black}, textfont={color=black}}
    \caption{Distribution of stars of the projects where AI-generated code was present}
    \captionsetup{labelfont={}, textfont={}}
    \label{fig:stars}
\end{figure}

\begin{figure}[htbp]
    \centering
    \includegraphics[width=0.99\textwidth]{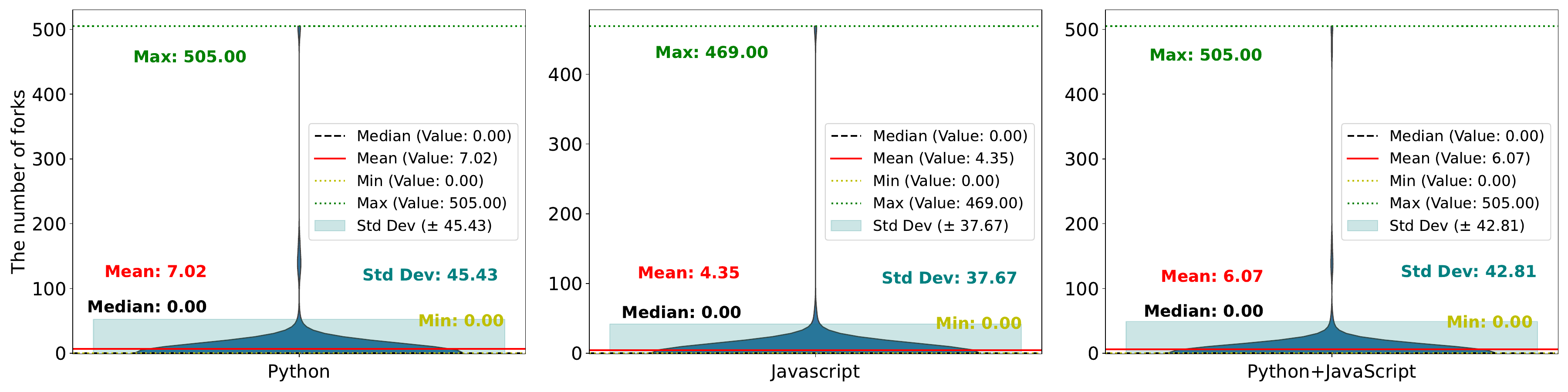}
    \captionsetup{labelfont={color=black}, textfont={color=black}}
    \caption{Distribution of forks of the projects where AI-generated code was present}
    \captionsetup{labelfont={}, textfont={}}
    \label{fig:forks}
\end{figure}

\textcolor{black}{\textbf{Step 3.} To understand the features of our dataset, we first conducted a statistical analysis of the stars and forks of the projects where AI-generated code was present. The distribution of stars and forks are shown in Figure~\ref{fig:stars} and Figure~\ref{fig:forks}, respectively. We found that the projects containing AI-generated code often tend to be small and have low popularity in the GitHub community, possibly often developed by individuals or small teams. This finding is consistent with the recent study by Yu \textit{et al.} about the characteristics of LLM-generated code on GitHub~\cite{yu2024large}. We speculated that the low popularity may be correlated with the age of the project. Most of these projects are relatively new as Copilot itself is also a recent development. Despite the low popularity and user engagement of most projects, there are a few exceptions in the dataset, such as the project \texttt{OnmyojiAutoScript}, which received 1,672 stars and 505 forks.}

\begin{figure}[htbp]
    \centering
    \includegraphics[width=0.99\textwidth]{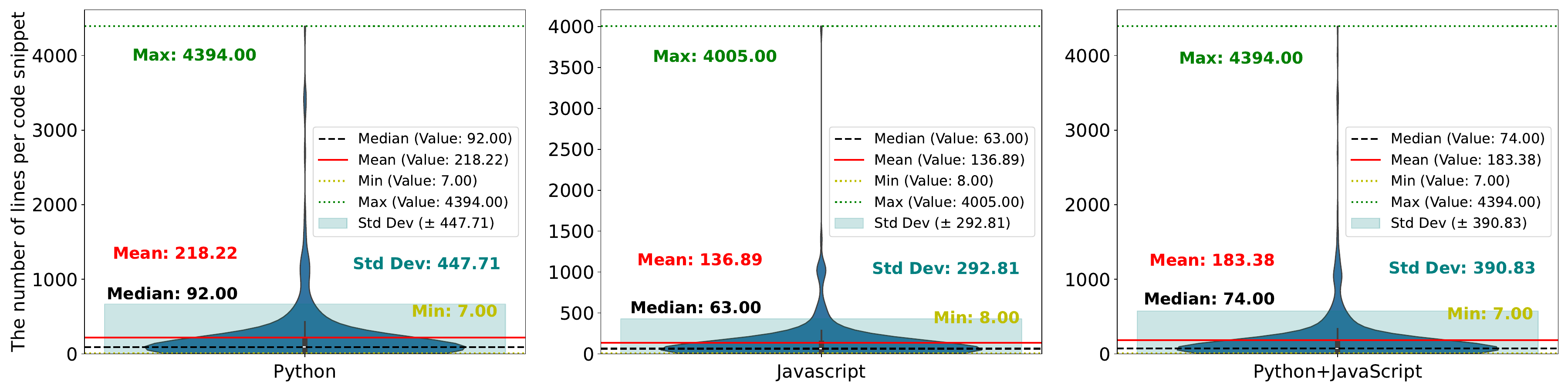}
    \captionsetup{labelfont={color=black}, textfont={color=black}}
    \caption{Distribution of LoC of the files where AI-generated code was present}
    \captionsetup{labelfont={}, textfont={}}
    \label{fig:lines}
\end{figure}

\textcolor{black}{In addition, we also counted the lines of code (LoC) of the collected code files. Figure~\ref{fig:lines} shows the distribution of LoC in different languages (Python and JavaScript respectively) and then the LoC of all code files (Python + JavaScript) in the dataset. We found that the median LoC of code is 74, while the average reaches 181. Although the files obtained from GitHub are relatively small in size, they are still significantly larger than the size of code generated using designed scenario prompts. For example, when Majdinasab \textit{et al.}~\cite{majdinasab2024assessing} replicated the experiment by Pearce \textit{et al.}~\cite{pearce2022asleep}, the median LoC of code generated by their designed prompts through Copilot was 42, with an average of 40. While using the SecurityEval dataset~\cite{siddiq2022securityeval} as prompts to test Copilot, the LoC of generated code is concentrated between 5 and 26. This indicates that our collected dataset has an advantage in terms of the size of the generated code.} 

\textcolor{black}{We also analyzed the application domains of the code snippets in our dataset. To understand the application domains, we classified the code in the dataset based on the project description and the specific functions of the code files. Two authors independently classified all code snippets within a period of ten days. If their categorization results diverged, the two authors discussed until they reached a consensus. In total, we categorized the generated code into six application domain categories: Game, Web Application, Utility Tool, AI Application, Network Communication, and Others. Table~\ref{tab:code_categories} shows the descriptions of application domains and the categorization of code snippets in different programming languages.}

\begin{table}[htbp]
\normalsize
\centering
\captionsetup{labelfont={color=black}, textfont={color=black}}
\caption{Application domains of the files where AI-generated code was present}
\captionsetup{labelfont={}, textfont={}}
\renewcommand{\arraystretch}{1.5} 
\resizebox{\textwidth}{!}{%
\begin{tabular}{|p{3cm}|p{8cm}|p{2cm}|p{2cm}|}
\hline
\textbf{Category} & \textbf{Description} & \textbf{Python } & \textbf{JavaScript} \\ \hline
Utility Tool & Involves scripts, command-line tools, and other programs that simplify daily tasks. & 39.0\% & 30.1\% \\ \hline
Web Application & Code related to front-end, back-end development, and server-side operations. & 15.6\% & 49.8\% \\ \hline
AI Application & Code for machine learning, deep learning, natural language processing, or other intelligent applications. & 15.6\% & 2.8\% \\ \hline
Game & Contains code related to game development, using engines such as Unity, Unreal, and Godot. & 15.9\% & 8.5\% \\ \hline
Network Communication & Involves code for implementing network protocols, client/server communication. & 8.2\% & 5.7\% \\ \hline
Others & Code that does not fit into the above categories. & 5.4\% & 2.8\% \\ \hline
\end{tabular}%
}
\label{tab:code_categories}
\end{table}


\subsection{Data Analysis and Results Filtering}\label{Data Analysis and Results Filtering}
\subsubsection{Data Analysis}
\textbf{Step 4.} \textcolor{black}{We used well-known automated static analysis tools suggested by OWASP~\cite{owasp} to scan the collected code snippets. The reason why we chose a static rather than dynamic approach is that the static analysis has higher coverage and is able to analyze programs without executing them~\cite{dunlap2023finding}. While dynamic analysis offers more precise insights by reasoning about program behavior, it suffers from limited coverage and can be expensive~\cite{sui2020recall,ernst2003static,tahir2012systematic}. Static security analysis tools have been widely used by developers and companies~\cite{distefano2019scaling,sadowski2018lessons}. 
Employing static analysis enables us to analyze program segments (smaller snippets) without requiring the entire program to be executed. 
Since different static analysis tools may use different algorithms and rules to detect security weaknesses, using multiple tools can increase our chances of discovering security weaknesses in the code. To improve the coverage and accuracy of the results, we used two static analysis tools for security checks on each code snippet (i.e., CodeQL plus one dedicated tool for the specific language, Bandit for Python, and ESLint for JavaScript).}

CodeQL is a scalable static security analysis tool that is widely used in practice and allows users to analyze code and detect relevant weaknesses using predefined queries and test suites and supports for multiple languages (including Java, JavaScript, C++, C\#, Go, and Python~\cite{codeql}). Before using CodeQL to scan the identified code snippets for security weaknesses, we needed to create a CodeQL database for the source code. Source code can be directly analyzed for interpreted languages like Python and JavaScript without being compiled into intermediate code. Thus, we directly use the command line to generate the database needed for CodeQL queries. Then, we used CodeQL to analyze the generated database. \textcolor{black}{We can find the available query suites for different programming languages in the \texttt{codeql} repository.} For example, the \texttt{codeql/python-queries} package contains the following query suites~\cite{codeqlcli}:

\begin{itemize}
    \item \texttt{Python-code-scanning.qls}, the standard scanning query for Python. It covers various features and syntax of Python and aims to discover some common weaknesses in the code.
    \item \texttt{Python-security-extended.qls}, which includes some more advanced queries than \texttt{Python-code-scanning.qls} and can detect more security weaknesses.
    \item \texttt{Python-security-and-quality.qls}, which combines queries related to security and quality, covering various aspects of Python development, from basic code structure and naming conventions to advanced security and performance weaknesses. It aims to help developers improve the security and quality of their code.
\end{itemize}

\textcolor{black}{In this study, we scanned code snippets using the \texttt{<language>-security-and-quality.qls} test suite. These test suites check for multiple security properties and cover many CWEs. For example, the \texttt{python-security-and-quality.qls} test suite for Python provides 168 security checks, and the JavaScript test suite provides 203 security checks. As the query reports only provide the name and description of the security weaknesses, we manually matched the results in the query reports with the corresponding CWE IDs based on the official documentation provided~\cite{codeqlhelp}.}

\textcolor{black}{For Python files, we used  Bandit~\cite{bandit} -- a Python's static security analysis tool. When running  Bandit, we enabled all security rules by default. Bandit had a total of 73 security check rules, and each Bandit rule was typically associated with a corresponding CWE number, helping developers better understand the categories of detected vulnerabilities and align with industry standards. For JavaScript files, we used ESLint~\cite{eslint} -- a widely used JavaScript static analysis tool that can detect common coding issues and security vulnerabilities. Since we needed to configure the security check rules ourselves, we used the \texttt{eslint-plugin-security} plugin for security checks, which is an ESLint plugin containing 14 security check rules designed to detect common security vulnerabilities in JavaScript. In addition, we also configured the plugins \texttt{eslint-plugin-no-secrets} and \texttt{eslint-plugin-xss} with more precise rules to identify security weaknesses more accurately. For example, some of the rules in \texttt{eslint-plugin-security} also deal with XSS protection (such as avoiding unsafe \texttt{eval()} usage), but \texttt{eslint-plugin-xss} provides more precise rules against front-end XSS attacks. Both tools accept raw code snippets as input, and there is no need to pre-process the data before using them.}
\textcolor{black}{It is worth mentioning that the detection rules of these static analysis tools do not have a one-to-one correspondence with CWEs. For example, multiple detection rules in Bandit may identify security issues that can all be classified under \textit{CWE-78: OS Command Injection}. As a result, we did not explicitly provide the detection rule-to-CWE correspondences. Instead, we manually mapped the detection rules to the CWEs after using these static analysis tools in conjunction with specific warning messages. We have provided the detection rule-to-CWE correspondences in our replication package~\cite{dataset}.}

As explained in Section \ref{Data Collection and Filtering}, we considered the code snippet from the \textit{Repository} label to be the entire code file, while the code snippet from the \textit{Code} label exists in the code file, with the exact location of the code snippets unspecified at this stage.


\subsubsection{Results Filtering}\label{Results Filtering}
\textbf{Step 5.} \textcolor{black}{We scanned code snippets from the \textit{Repository} and \textit{Code} labels, and we filtered the analysis results before analyzing them. In this step, we adopted part of the strategies used by developers when they perform different tasks with static analysis tools: \textit{warning prioritization} and \textit{determining whether a warning is a false positive or a true report}~\cite{do2020software}. We first performed an initial filtering of the results based on the priority of the warnings. Specifically, we manually removed the \textcolor{black}{duplicate} analysis results that were reported by the two tools (14 duplicate results by Bandit and 2 duplicate results by ESLint), keeping only one for further analysis. The first author eliminated analysis results that were not security weaknesses. We identified three types of warnings from CodeQL analysis:}

\textcolor{black}{\begin{itemize}
    \item \textit{Recommendation}, which provides suggestions for improving code quality;
    \item \textit{Warning}, which alerts to potential weaknesses that could cause code to run abnormally or unsafely;
    \item \textit{Error}, which is the highest level of warning and alert to inform that the error could cause code to fail to compile or run incorrectly.
\end{itemize}}

\textcolor{black}{Since our research primarily focused on security weaknesses, we only counted code snippets that had \textit{warnings} and \textit{errors}, and we ignored \textit{recommendations} on code quality.} 

\textcolor{black}{After obtaining preliminary analysis results related to security weaknesses, we manually checked the analysis results to confirm that the security weaknesses were actually caused by the automatically generated code.} For the initial results obtained under the \textit{Repository} label, we checked them one by one. Specifically, we determined whether the corresponding location of the code snippet indeed had a security weakness based on the line number information provided by the analysis results. For the initial results obtained under the \textit{Code} label, we confirmed whether the security weaknesses truly existed and verified if the security weaknesses were caused by Copilot-generated code. Specifically, after scanning the code file, we pinpointed the code snippet within the file based on the line number of the security weakness indicated in the analysis results. We assessed whether it was generated by Copilot or other generation tools by checking the surrounding comments and determined whether a security weakness existed in that particular context. We further analyzed the filtered analysis results in \textbf{Step 6}, detailed in Section~\ref{Results}, according to the specific RQs.

\textcolor{black}{\subsection{Fixing Code with Security Weaknesses}\label{security weakness Mitigation} \textbf{Step 7.} Based on RQ2 results, we randomly selected a subset of code snippets containing security weaknesses as the repair dataset since our purpose is to explore the feasibility of \textit{Copilot Chat} in fixing the security weaknesses. These code snippets were all generated by Copilot, as most of the code obtained from GitHub was produced by Copilot. Since we aimed to fix as many security weaknesses as possible, we did not use clean data (i.e., code snippets without known security weaknesses). We randomly selected 50 Python and 40 JavaScript code snippets from the \textit{Repo} label, which contained 163 and 132 security weaknesses, respectively. We utilized \textit{Copilot Chat} to remediate the security weaknesses identified by the static analysis tools. We conducted several experiments on the selected dataset with 295 security weaknesses using three different prompts. \textit{Copilot Chat} provides developers with basic slash commands, avoiding the need to write complex prompts for common scenarios. The slash commands include: \texttt{/tests:} generates unit tests for the selected code. \texttt{/fix:} provides fix suggestions for issues in the selected code. \texttt{/explain:} explains the selected code. \texttt{/optimize:} analyzes and improves the runtime of the selected code. We first used the slash command \texttt{/fix} to provide fix suggestions for the selected code. 
Then, we performed the fixes using our designed basic prompts and enhanced prompts with additional context. The prompt designs are as follows:}
\begin{tcolorbox}[colback=gray!10, colframe=gray!80, title=Basic Prompt]
\textit{Is there a security weakness with this code snippet? If so, please fix it and return the fixed code.}
\end{tcolorbox}

\begin{tcolorbox}[colback=gray!10, colframe=gray!80, title=Enhanced Prompt]
\textit{This code snippet is known to have some security weaknesses. The warning messages are as follows: \texttt{<insert warning messages here>}. Please fix these security weaknesses and return the fixed code.}
\end{tcolorbox}

\begin{figure*}[htbp]
	\centering
	\includegraphics[ width=0.9\textwidth]{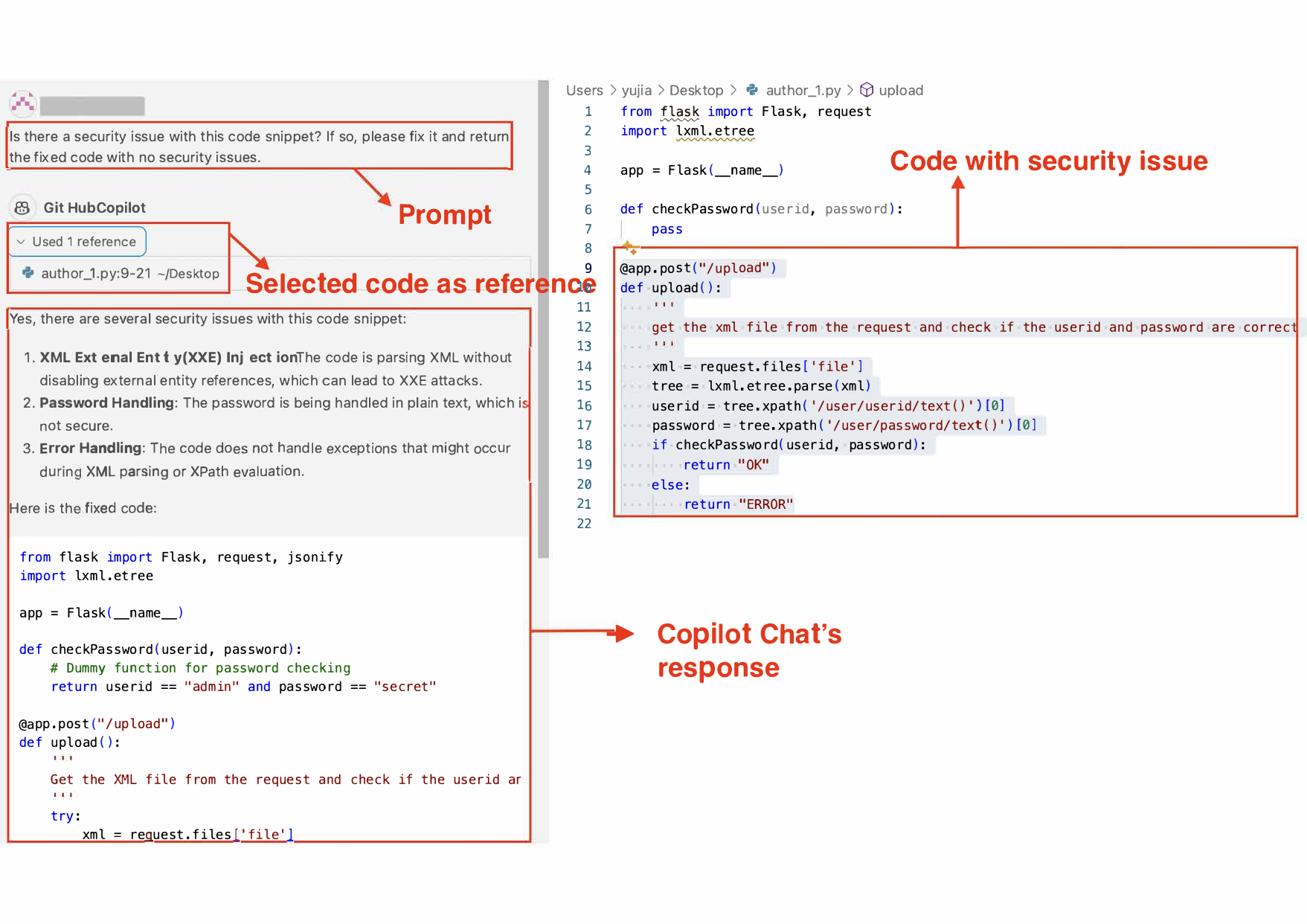}
    \captionsetup{labelfont={color=black}, textfont={color=black}}
	\caption{An example of using \textit{Copilot Chat} to fix security weaknesses in Visual Studio Code}
    \captionsetup{labelfont={}, textfont={}}
	\label{fig:Example of copilot chat}
\end{figure*}
\textcolor{black}{Figure~\ref{fig:Example of copilot chat} shows an example of how to use \textit{Copilot Chat} to fix the security weaknesses in the code. We first select code snippets with security weaknesses and provide the three prompts (\texttt{/fix}, basic prompt, and enhanced prompt) in the IDE chat window. Then, \textit{Copilot Chat} uses the selected code as a reference (input) to generate responses to our prompts, which we can accept or reject. 
Considering the token limitation issue, we always selected the function block containing the security weaknesses as the reference for \textit{Copilot Chat}. After accepting the fix suggestions from \textit{Copilot Chat}, we used the static analysis tools to recheck the modified code snippets. We recorded the positions of the modified code, indicating where the accepted code is located. Then we examined whether security weaknesses still exist in the corresponding positions of the modified code according to the rechecking results of the static analysis tools.}

\textcolor{black}{We provided the complete dataset (including code snippets, full analysis results, and filtered results) in our replication package~\cite{dataset}.}

\section{Results}\label{Results}

In this section, We present the results of the three RQs formulated in Section \ref{Goal and Research Questions}. For each RQ, we first explain how we analyzed the collected code snippets to answer the RQ, and we then answer each of our three RQs.
\subsection{RQ1: How secure is the code generated by Copilot in GitHub projects?}\label{RQ1}

\noindent\textbf{Approach}. To answer this RQ, we collected 733 Python and JavaScript code snippets generated by Copilot and two other tools from GitHub projects, with 672 from Copilot, 38 from CodeWhisperer, and 23 from Codeium. We used two static analysis tools (CodeQL + another language-dedicated tool, Bandit for Python, and ESLint for JavaScript) to scan. The aim is to achieve better coverage of security weaknesses. Therefore, as long as one of the tools detected the presence of a security weakness, the code snippet was considered vulnerable.

\textcolor{black}{Subsequently, we filtered the analysis results as described in Section~\ref{Results Filtering}.} For the analysis results from the \textit{Repository} label, we marked them as 1 (security weakness exists) and 0 (no security weakness exists). For the analysis results from \textit{Code} label, we marked them as 1 (security weakness exists) and 0 (no security weakness exists or the code snippet with a security weakness was not generated by code generation tools). 
\begin{figure*}[htbp]
	\centering
	\includegraphics[width=0.9\linewidth]{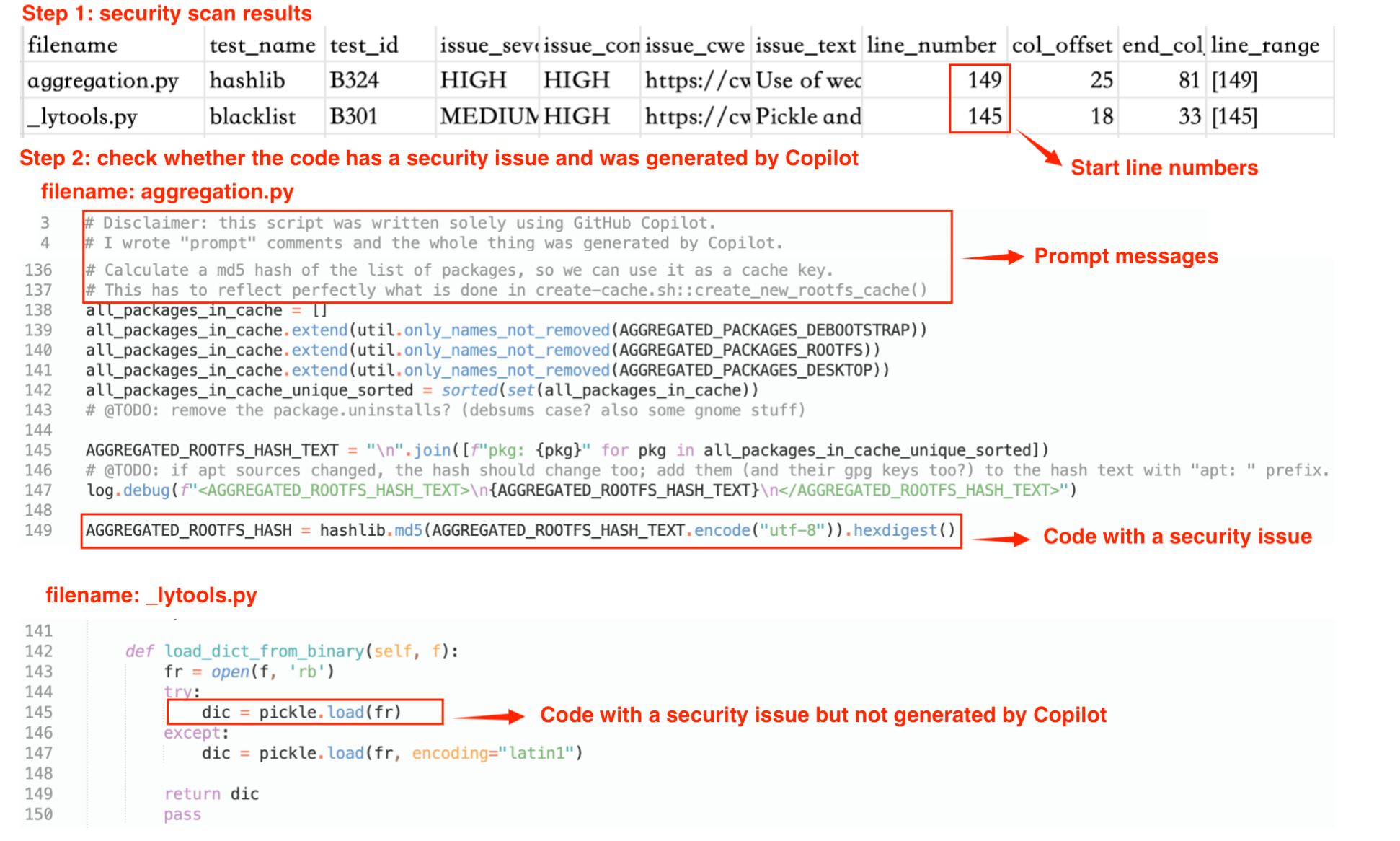}
	\caption{Example of filtering analysis results from the \textit{Code} label that are generated by Copilot}
	\label{fig:filtering scan result}
\end{figure*}
When manually checking the analysis results from the \textit{Code} label, we also needed to identify whether the security weaknesses obtained from the scan were from automatically generated code snippets based on the comment (prompt message) that appears before the method. We provide a working example of the filtration of the analysis results in Figure~\ref{fig:filtering scan result}. In \textbf{Step 1}, we first went to the corresponding file to locate the specific code snippet based on the start line numbers of the analysis results. In \textbf{Step 2}, we located the code at Line 149 in \texttt{aggregation.py}. We found that this code does have a security weakness and determined that it was generated by Copilot based on the prompt messages above.
Consequently, we marked the corresponding security analysis result as ``1''. We also located the code at Line 145 in \texttt{\_lytools.py}. We found that this code had a security weakness, but there were no prompt messages that would indicate that it was generated by Copilot. Consequently, we marked the corresponding security analysis result as ``0'' and discarded this analysis result from further analysis. 

\textcolor{black}{Two authors independently filtered the analysis results within a period of two weeks, and if there were any results that they were unsure of or disagreed with, the two authors discussed the results until they reached an agreement. Cohen’s Kappa coefficients~\cite{Jacob1960Coefficient} were 0.85 for the analysis results from the \textit{Repository} label, and 0.82 for the analysis results from the \textit{Code} label, which were both higher than 0.8, ensuring a good accuracy of the filtering results. Finally, we kept the results marked as 1 and aggregated the filtered results obtained using multiple analysis tools to calculate the number of code snippets with security weaknesses detected. All labeling results can be found in our online replication package~\cite{dataset}.}

\begin{figure}
    \begin{subfigure}{0.45\textwidth}
        \includegraphics[width=\linewidth]{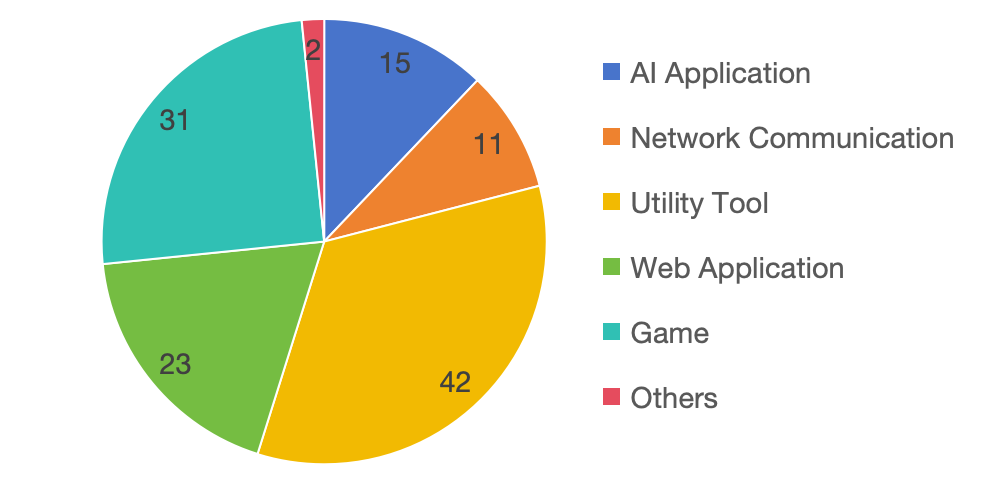}
        \caption{application domains of Python code}
        \label{fig:domain-py}
    \end{subfigure}
    \hfill
    \begin{subfigure}{0.45\textwidth}
        \includegraphics[width=\linewidth]{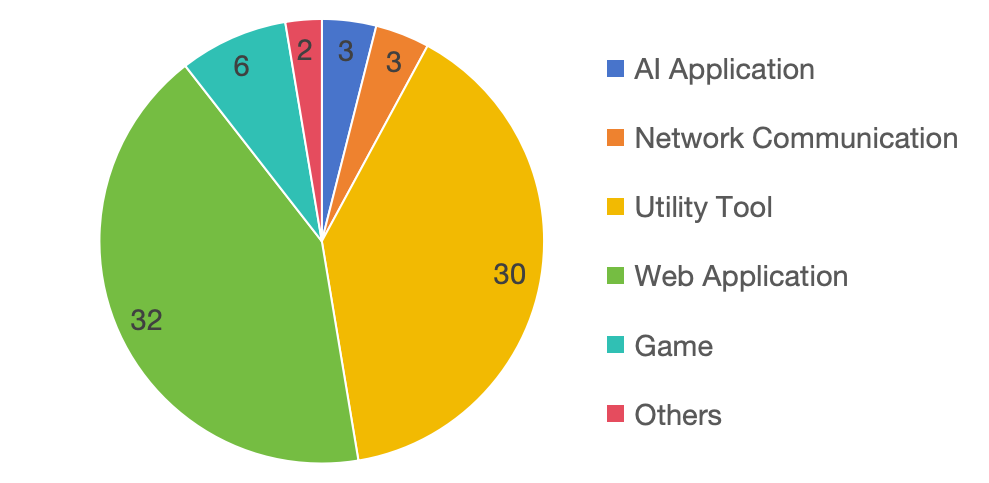}
        \caption{application domains of JavaScript code}
        \label{fig:domain-js}
    \end{subfigure}
    \captionsetup{labelfont={color=black}, textfont={color=black}}
    \caption{Application domains of the code snippets with security weaknesses}
    \captionsetup{labelfont={}, textfont={}}
    \label{fig:domain}
\end{figure}
\begin{figure}[htbp]
    \centering
    \includegraphics[width=0.99\textwidth]{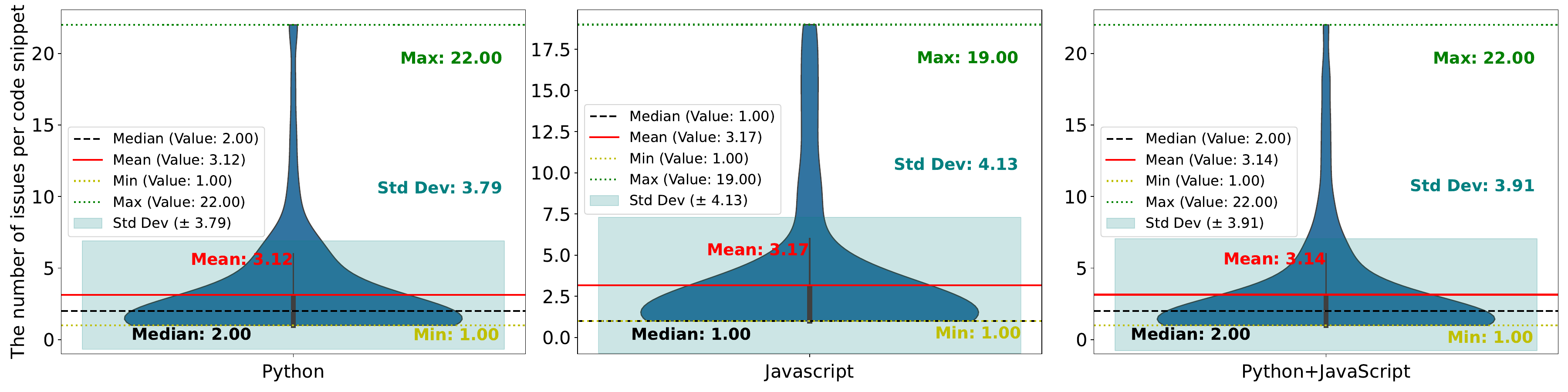}
    \captionsetup{labelfont={color=black}, textfont={color=black}}
    \caption{Distribution of the number of security weaknesses per code snippet}
    \captionsetup{labelfont={}, textfont={}}
    \label{fig:numbers}
\end{figure}

\noindent\textcolor{black}{\textbf{Results}. Table \ref{The number and percentage of code snippets with security weaknesses} shows the number of code snippets containing security weaknesses and the number of security weaknesses identified in the code snippets. From the statistical results, we found that out of the 733 generated code snippets, 27.3\% of them contained security weaknesses. Around 30\% of Python code snippets (124 of the 419) and around 25\% of JavaScript code snippets (76 of the 314) contained security weaknesses. At the same time, we can see that more than half of the code snippets containing security weaknesses (102 of 200) have more than one security issue. In Figure~\ref{fig:domain}, we present the distribution of application domains for the 200 code snippets (27.3\% of all) identified with security weaknesses.
We also found 628 security weaknesses in the 200 code snippets. Note that one code snippet may contain multiple security weaknesses attributed to multiple CWEs. Figure~\ref{fig:numbers} shows the distribution of security weaknesses per code snippet. Our analysis reveals an average incidence of 3 security weaknesses per code snippet, irrespective of the programming language. The maximum count of security weaknesses within a single code snippet, which consisted of 1,595 LoC, reached 22.}

\begin{table}[htbp]
\footnotesize
\centering
\captionsetup{labelfont={color=black}, textfont={color=black}}
\caption{Statistics of the code snippets with security weaknesses}
\captionsetup{labelfont={}, textfont={}}
\label{The number and percentage of code snippets with security weaknesses}
{
\begin{tabular}{lcccc}
\toprule
\textbf{Language}   & \textbf{\#  Snippets}    & \textbf{\# Security weaknesses}               &\textbf{\begin{tabular}[c]{@{}l@{}}\# Snippets containing \\ security weaknesses\end{tabular} }   & \textbf{\begin{tabular}[c]{@{}l@{}}\# containing more than \\ one security weakness\end{tabular} }\\ \midrule
 Python        & 419  & 387   & 124 (29.5\%)  & 65 (15.5\%) \\ 
 JavaScript    & 314  & 241   & 76 (24.2\%)   & 37 (11.8\%) \\  \midrule
 \textbf{Total}        &  \textbf{ 733 } & \textbf{628} &      \textbf{ 200 (27.3\%) }   & \textbf{ 102 (13.9\%) }\\ \bottomrule
\end{tabular}
}
\end{table}

\subsection{RQ2: What security weaknesses are present in the code generated by Copilot across different application domains?}\label{RQ2}

\noindent\textbf{Approach}. To answer RQ2, we processed the analysis results collected by RQ1, which were manually checked to see if they contained security weaknesses. For each code snippet, we used CWEs to classify the security weaknesses identified by the static analysis tools. Each CWE has a unique ID and a set of related descriptions, including its potential impact and how to detect and fix the CWE~\cite{cwe}.
\textcolor{black}{To accurately identify the type of security weakness with the corresponding CWE, we did not directly accept the information provided by the static analysis tools. Instead, we manually identified the type of security weakness at the specific location reported in the warning message of the static analysis tool. \textcolor{black}{Only Bandit provided the CWE IDs of the security issues identified, and we did not follow the CWE IDs of 55 out of 235 security issues reported by Bandit. For example, Bandit reported ``\textit{Starting a process with a partial executable path}'' as belonging to \textit{CWE-78: OS Command Injection}. After our manual assessment, we concluded that it was more appropriate to map this warning message to \textit{CWE-427: Uncontrolled Search Path Element}, as the security issue was primarily related to path control rather than command construction.}} We utilized the warning information provided by static analysis tools, such as the test\_name and message fields indicated by CodeQL, in conjunction with relevant descriptions from the CWE, to perform manual correspondences. \textcolor{black}{Initially, two authors independently matched each description of the security issue with a CWE ID within a period of ten days. Cohen's Kappa coefficient~\cite{Jacob1960Coefficient} was 0.82, which was higher than 0.8 and indicated a high level of agreement between the two authors. In case of disagreement, a discussion was initiated between the two authors, and one other author (a security expert) was then involved to provide his assessment.} We provided correspondences between the warning prompt messages and CWEs in our online replication package~\cite{dataset}.
In the final stage, we performed a statistical analysis of CWE weaknesses in 200 code snippets that contained security weaknesses. Additionally, we analyzed the distribution of different CWEs across different application domains of the code snippets.

We also discussed whether the security weaknesses present in the code generated by Copilot are widely prevalent in the code produced during software development. The code snippets in our collected dataset were mainly generated in 2022 and 2023. To compare whether the security weaknesses in in our dataset are widespread in this period, we chose MITRE 2023 CWE Top-25 list~\cite{cwe2023} as our baseline. We compared the CWEs obtained in RQ2 with the 2023 CWE Top 25.

\footnotesize
\begin{longtable}{m{0.1\linewidth} m{0.5\linewidth} m{0.15\linewidth}<{\centering} m{0.1\linewidth}<{\centering}}
\captionsetup{labelfont={color=black}, textfont={color=black}}
\caption{Distribution of CWEs in code snippets} \label{Distribution of CWEs in code snippets} \\
\captionsetup{labelfont={}, textfont={}}
\hline
\textbf{CWE-ID}  & \textbf{Name} & \textbf{Frequency} & \textbf{Percentage} \\ \hline
\endfirsthead
\multicolumn{4}{c}%
{{\tablename\ \thetable{} -- continued from previous page}} \\
\hline
\textbf{CWE-ID}  & \textbf{Description} & \textbf{Frequency of Specific CWE} & \textbf{Percentage} \\ \hline
\endhead
\hline \multicolumn{4}{r}{{Continued on next page}} \\
\endfoot
\hline
\endlastfoot
CWE-330 & Use of Insufficiently Random Values Weakness & 114 & 18.15\% \\ 
\textbf{CWE-94}  & Improper Control of Generation of Code ('Code Injection') & 62 & 9.87\% \\ 
\textbf{CWE-79}  & Improper Neutralization of Input During Web Page Generation ('Cross-site Scripting') & 60 & 9.55\% \\ 
\textbf{CWE-78}  & Improper Neutralization of Special Elements used in an OS Command ('OS Command Injection') & 39 & 6.21\% \\ 
CWE-427 & Uncontrolled Search Path Element & 35 & 5.57\% \\ 
CWE-457 & Use of Uninitialized Variable & 30 & 4.78\% \\ 
\textbf{CWE-22}  & Improper Limitation of a Pathname to a Restricted Directory ('Path Traversal') & 29 & 4.62\% \\ 
CWE-772 & Missing Release of Resource after Effective Lifetime & 29 & 4.62\% \\ 
\textbf{CWE-89} & Improper Neutralization of Special Elements used in an SQL Command ('SQL Injection') & 27 & 4.30\% \\ 
CWE-259 & Use of Hard-coded Password & 16 & 2.55\% \\ 
CWE-685 & Function Call with Incorrect Number of Arguments & 14 & 2.23\% \\ 
CWE-284 & Improper Access Control & 13 & 2.07\% \\ 
CWE-312 & Cleartext Storage of Sensitive Information & 12 & 1.91\% \\ 
CWE-390 & Detection of Error Condition Without Action & 11 & 1.75\% \\ 
CWE-456 & Missing Initialization of Variable & 11 & 1.75\% \\ 
CWE-607 & Use of Wrong Operator in String Comparison & 11 & 1.75\% \\ 
CWE-770 & Allocation of Resources Without Limits or Throttling & 11 & 1.75\% \\ 
\textbf{CWE-20} & Improper Input Validation& 8 & 1.27\%\\
CWE-665 & Improper Initialization & 8 & 1.27\%\\
CWE-117 & Improper Output Neutralization for Logs & 7 & 1.11\% \\
CWE-400 & Uncontrolled Resource Consumption & 7 & 1.11\% \\
CWE-561 & Dead Code & 7 & 1.11\% \\ 
CWE-732 & Incorrect Permission Assignment for Critical Resource & 6 & 0.96\% \\ 
\textbf{CWE-798} & Use of Hard-coded Credentials & 6 & 0.96\% \\ 
CWE-215 & Information Exposure Through Debug Information & 5 & 0.80\% \\ 
CWE-290 & Authentication Bypass by Spoofing & 5 & 0.80\% \\ 
CWE-295 & Improper Certificate Validation & 5 & 0.80\% \\ 
CWE-209 & Information Exposure Through an Error Message & 4 & 0.64\% \\ 
CWE-252 & Unchecked Return Value & 4 & 0.64\% \\ 
CWE-571 & Expression is Always True & 4 & 0.64\% \\ 
CWE-605 & Multiple Binds to the Same Port & 4 & 0.64\% \\ 
CWE-200 & Information Exposure & 3 & 0.48\% \\ 
CWE-327 & Use of a Broken or Risky Cryptographic Algorithm & 3 & 0.48\% \\ 
CWE-367 & Time-of-check Time-of-use Race Condition & 3 & 0.48\% \\ 
CWE-563 & Assignment of a Fixed Address to a Pointer & 3 & 0.48\% \\ 
CWE-116 & Improper Encoding or Escaping of Output & 2 & 0.32\% \\ 
CWE-480 & Use of Incorrect Operator & 2 & 0.32\% \\ 
CWE-502 & Deserialization of Untrusted Data & 2 & 0.32\% \\ 
CWE-601 & URL Redirection to Untrusted Site ('Open Redirect') & 2 & 0.32\% \\ 
CWE-208 & Observable Timing Discrepancy & 1 & 0.16\% \\ 
CWE-570 & Expression is Always False & 1 & 0.16\% \\ 
CWE-628 & Function Not Implemented Correctly & 1 & 0.16\% \\ 
CWE-682 & Incorrect Calculation & 1 & 0.16\% \\  \hline
 
\textbf{43 Types}      &     &  \textbf{Total: 628} &\\ \hline
\end{longtable}

\normalsize
\noindent\textcolor{black}{\textbf{Results}. Table \ref{Distribution of CWEs in code snippets} shows the distribution of CWEs in the code snippets and the total number of occurrences (frequency) of the CWEs in the code snippets. 
In total, we found 628 CWEs in 200 code snippets with security weaknesses. These security weaknesses were related to 43 CWE types, indicating that developers face a variety of security weaknesses when using Copilot and other code generation tools. \textit{CWE-330: Use of Insufficiently Random Values} is the most frequently occurring CWE, as it represents 18.15\% of the security weaknesses, followed by \textit{CWE-94: Code Injection }, \textit{CWE-79: Cross-site Scripting}, and \textit{CWE-78: OS Command Injection}. Some CWEs appear less frequently. For example, \textit{CWE-117: Improper Output Neutralization for Logs} only occurred twice. Furthermore, many CWEs occur with a frequency of less than 1\%, for example, \textit{CWE-117: Improper Output Neutralization for Logs}and \textit{CWE-290: Authentication Bypass by Spoofing}. This indicates that the types of security weaknesses are closely related to the specific scenarios in which developers use code generation tools, emphasizing the importance of maintaining vigilance and caution when programming.
In addition, we highlighted those CWEs that are the ``Top-25 CWEs'' in \textbf{bold}. It is worth noting that 233 security weaknesses in the code snippet correspond to these eight CWEs, which belong to the Top-25 CWE list, indicating that the CWE Top-25 weaknesses are also prevalent in the code generated by Copilot and other tools. 
At the same time, we can see that \textit{CWE-79: Cross-site Scripting} and \textit{CWE-78: OS Command Injection} are among the most frequently occurring security weaknesses in our results and rank high on the Top-25 CWE list.} 

\textcolor{black}{Table~\ref{tab:cwe_types} presents the top 5 CWEs that appear in Python and JavaScript code snippets. We found that the detected CWEs vary across programming languages. CWEs in Python are mainly related to data processing and system calls, such as \textit{CWE-78: OS Command Injection} and \textit{CWE-427: Uncontrolled Search Path Element}. In contrast, CWEs in JavaScript are primarily associated with dynamic code generation problems and security issues in Web development, such as \textit{CWE-94: Code Injection} and \textit{CWE-79: Cross-site Scripting}.}

\begin{figure*}[htbp]
	\centering
	\includegraphics[width=0.99\linewidth]{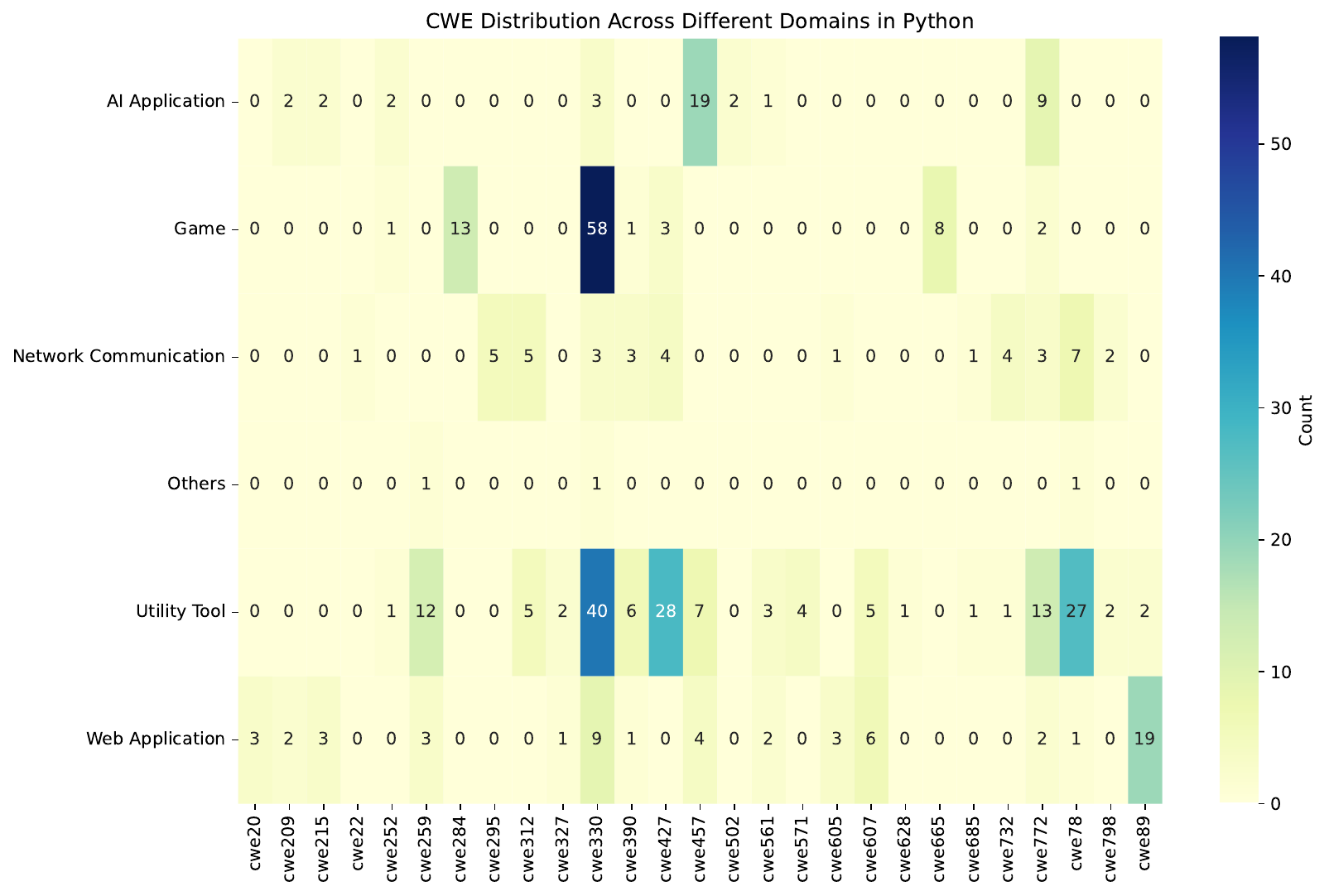}
    \captionsetup{labelfont={color=black}, textfont={color=black}}
	\caption{Distribution of CWEs in Python code snippets across different application domains}
    \captionsetup{labelfont={}, textfont={}}
	\label{fig:cwe-py}
\end{figure*}

\begin{figure*}[htbp]
	\centering
	\includegraphics[width=0.99\linewidth]{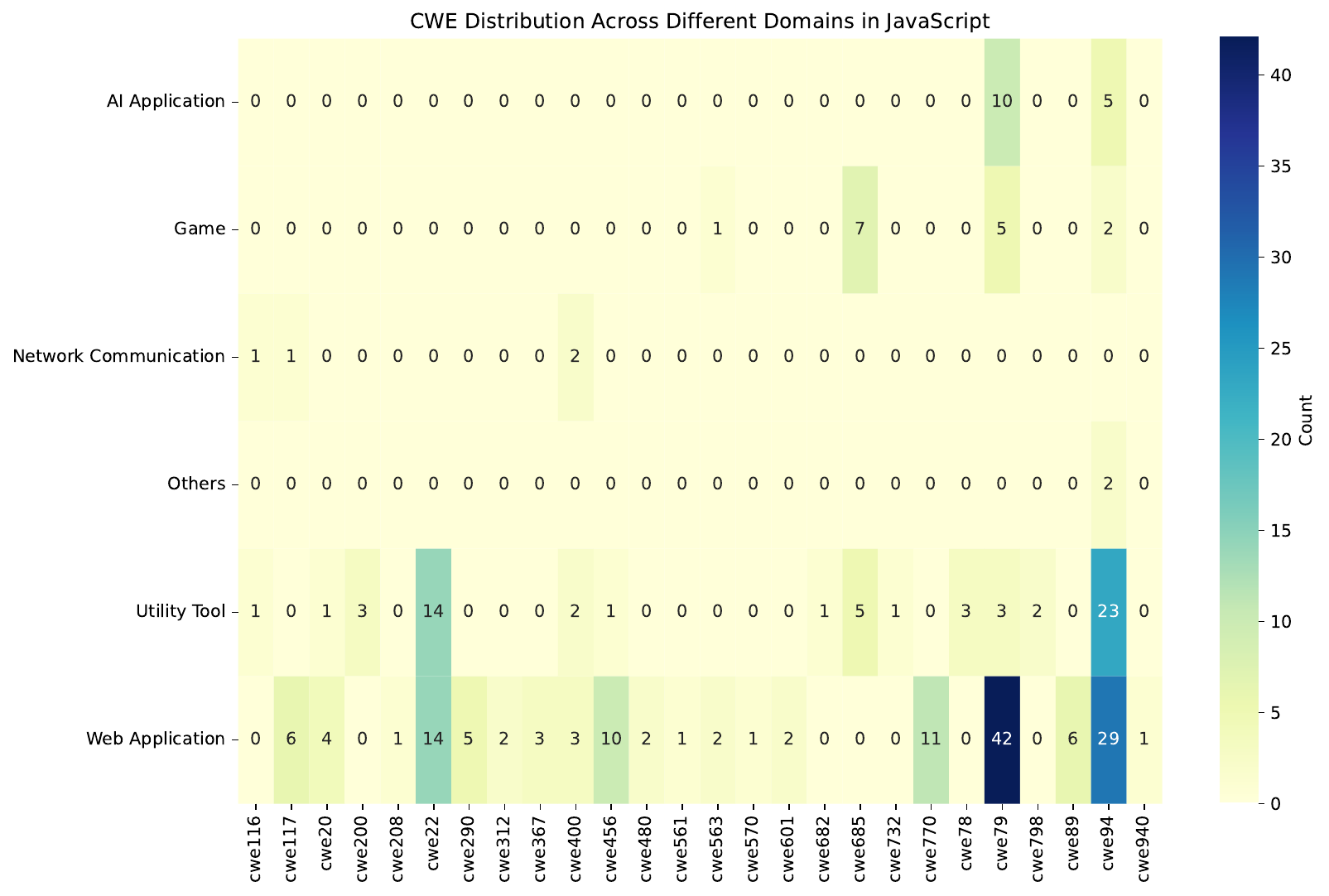}
    \captionsetup{labelfont={color=black}, textfont={color=black}}
	\caption{Distribution of CWEs in JavaScript code snippets across different application domains}
    \captionsetup{labelfont={}, textfont={}}
	\label{fig:cwe-js}
\end{figure*}

\textcolor{black}{Furthermore, Figure~\ref{fig:cwe-py} and Figure~\ref{fig:cwe-js} show the distribution of CWEs across different application domains. We can observe that the number and type of instances of CWE vary from one application domain to another. In addition to this, the distribution of CWEs within the same application domain varies by programming language. For example, in Python, the application domain with the most CWEs is Utility Tool, while in JavaScript, Web Applications have the most CWEs. Meanwhile, we can see that in the Web Application category, \textit{CWE-79: Cross-site Scripting} is the most frequent in JavaScript code snippets, while \textit{CWE-89: SQL Injection} is the most present in Python code snippets.}

\begin{table}[htbp]
\footnotesize
\centering
\caption{Top 5 CWEs in Python and JavaScript}
\label{tab:cwe_types}
\begin{tabular}{|c|p{5cm}|p{5cm}|}
\hline
\multirow{2}{*}{\textbf{Rank}} & \multicolumn{2}{c|}{\textbf{CWE Type}} \\ \cline{2-3} 
                                & \textbf{Python}          & \textbf{JavaScript}        \\ \hline
1 & CWE-330 (Use of Insufficiently Random Values Weakness) & CWE-94 (Improper Control of Generation of Code) \\ \hline
2 & CWE-78 (Improper Neutralization of Special Elements used in an OS Command) & CWE-79 (Improper Neutralization of Directives in Dynamically Evaluated Code) \\ \hline
3 & CWE-427(Uncontrolled Search Path Element) & CWE-22 (Improper Limitation of a Pathname to a Restricted Directory ('Path Traversal'))\\ \hline
4 & CWE-457 (Use of Uninitialized Variable) & CWE-685 (Function Call with Incorrect Number of Arguments) \\ \hline
5 & CWE-772 (Missing Release of Resource after Effective Lifetime) & CWE-770 (Allocation of Resources Without Limits or Throttling)\\ \hline
\end{tabular}
\end{table}
\normalsize

\textcolor{black}{\subsection{RQ3: Can Copilot Chat fix the security weaknesses of code generated by Copilot?}\label{RQ3}
\noindent\textbf{Approach}. To answer RQ3, we conducted experiments using \textit{Copilot Chat} on the randomly selected code snippets containing security weaknesses and asked \textit{Copilot Chat} to provide fix suggestions based on the three prompts (see Section~\ref{security weakness Mitigation}). We then fixed the code according to the suggestion provided by \textit{Copilot Chat}. Specifically, we used \textit{Copilot Chat} to repair snippets with security weaknesses. The repaired code snippets are then checked for security weaknesses using the same static analysis tools. After getting the analysis results, The first author used the same filtering steps described in Section~\ref{Results Filtering} to filter the results. We first removed the results that were \textcolor{black}{repetitively} reported by two of the tools. Then, we removed the analysis results that were not security weaknesses. Among the three types of warnings from CodeQL analysis, we only counted code snippets
that had \textit{warnings} and \textit{errors}, and ignored \textit{recommendations} on code quality. Since the code snippets we selected all come from the \textit{Repo} label, it is easier to filter the results related to security weaknesses, as we did not need to locate the specific line numbers of the code generated by Copilot in each file. Based on the results of the two successive security analysis, we determine whether the security weakness was successfully removed and the code is fixed. }

\begin{figure*}[htbp]
	\centering
	\includegraphics[width=0.85\linewidth]{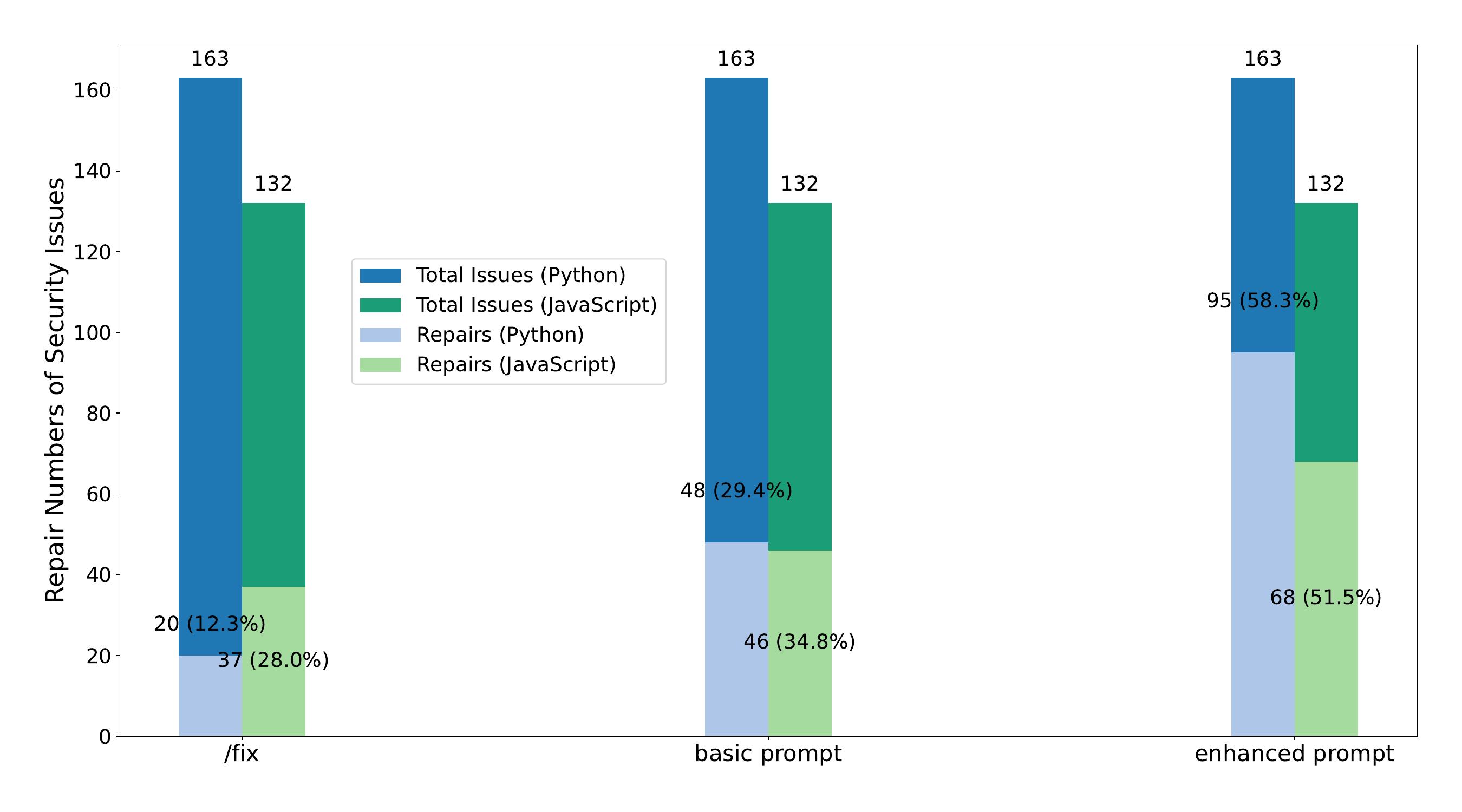}
    \captionsetup{labelfont={color=black}, textfont={color=black}}
	\caption{Fix numbers and percentages for security weaknesses by prompt and language}
    \captionsetup{labelfont={}, textfont={}}
	\label{fig:cwe-repair}
\end{figure*}

\noindent\textcolor{black}{\textbf{Results}. Table~\ref{table:fix} shows the proportion of CWEs that were fixed under the three prompts. We can see that using the slash command \/fix can fix 19.3\% of security weaknesses, while our designed basic and enhanced prompts can fix 31.8\% and 55.5\% of security weaknesses, respectively. \textit{Copilot Chat} demonstrates promising capability in fixing security weaknesses regardless of the prompt used. At the same time, providing enhanced information in the prompts helps improve the proportion of fixes. Figure~\ref{fig:cwe-repair} shows the number of security weaknesses fixed in code snippets of different languages when using various prompts. With basic prompts, security weaknesses were fixed in 29.4\% (48 of 163) of Python code and 34.8\% (46 of 132) in JavaScript. When using the enhanced prompt, the percentage of security weaknesses fixed in both languages rose to 58.3\% (95 of 163) and 51.5\% (68 of 132), respectively. We can see that \textit{Copilot Chat}'s ability to fix security weaknesses in Python and JavaScript snippets does not differ significantly.}

\begin{table}[h!]
\footnotesize
    \centering
    \captionsetup{labelfont={color=black}, textfont={color=black}}
    \caption{CWE issue fix percentages by the three prompts}
    \captionsetup{labelfont={}, textfont={}}
    \label{table:fix}
    \begin{tabular}{lccc}
        \hline
        \textbf{Prompt} & \textbf{Total CWE Before Fix}& \textbf{Total CWE After Fix} & \textbf{Fix Rate (\%)} \\ \hline
        /fix & 295 & 238 &    19.3\%\\ 
        basic & 295 & 201 &    31.8\%\\ 
        enhanced & 295 & 132 &  55.5\% \\ 
        \hline
    \end{tabular}
\end{table}

\begin{figure*}[htbp]
	\centering
	\includegraphics[width=\linewidth]{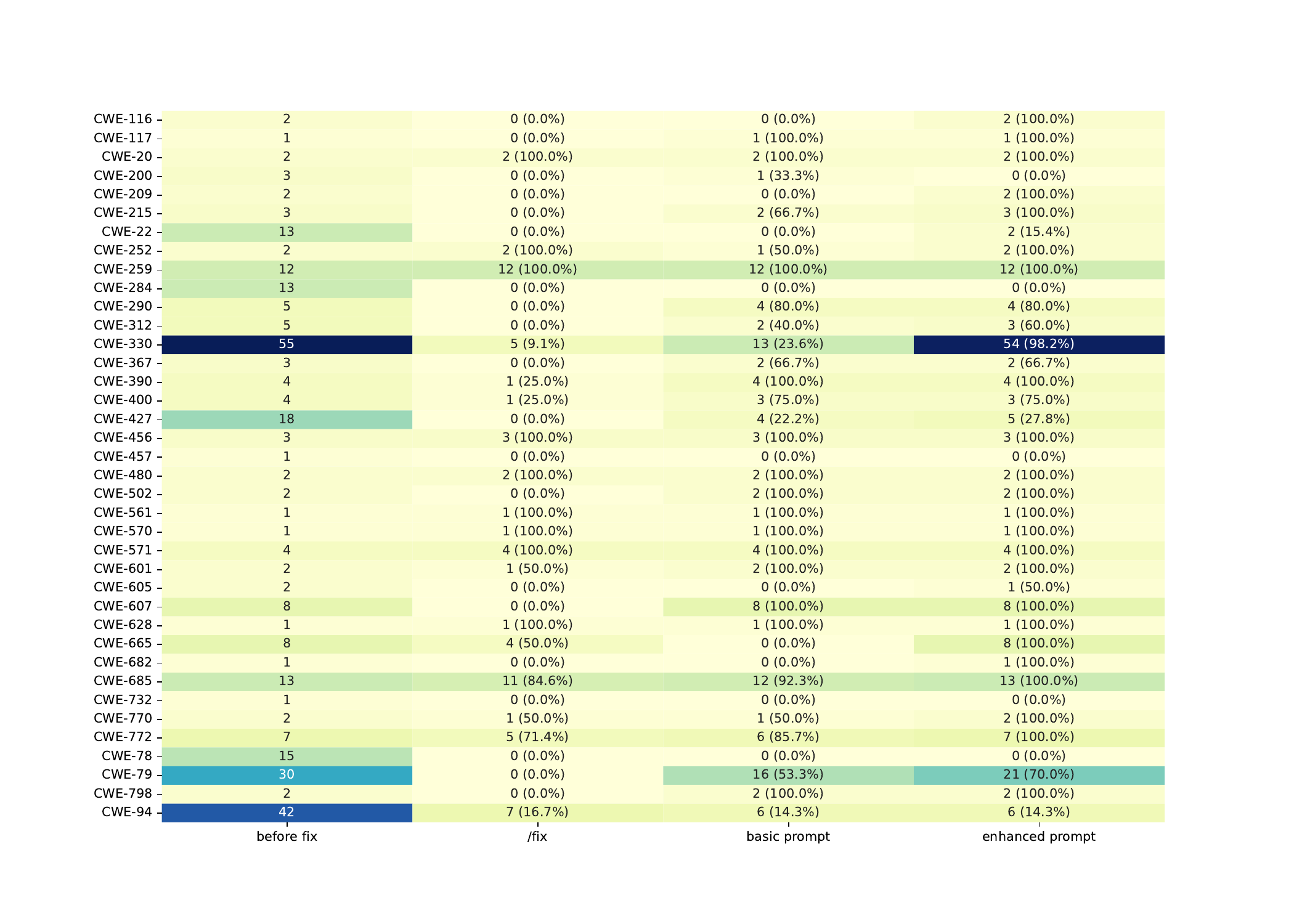}
        \captionsetup{labelfont={color=black}, textfont={color=black}}
	\caption{Distribution of the numbers of fixed CWEs under different prompts}
    \captionsetup{labelfont={}, textfont={}}
	\label{fig:cwe-fix}
\end{figure*}

\textcolor{black}{Furthermore, Figure~\ref{fig:cwe-fix} shows a heatmap with the number of different CWEs that were fixed when using different prompts. We can visually observe that \textit{Copilot Chat}'s effectiveness in fixing different CWEs varies. For certain CWEs, \textit{Copilot Chat} can perform well in repairing those snippets and provide a complete fix  (i.e., remove the CWE from the code) such as \textit{CWE-259: Use of Hard-coded Password} and \textit{CWE-685: Function Call with Incorrect Number of Arguments}, which can be 100\% fixed with the enhanced prompt. However, for some security weaknesses, the repair performance of \textit{Copilot Chat} is considerably poor as it fails to aviod certain CWEs. For example, for \textit{CWE-94: Code Injection}, Copilot Chat's fixes were successful in less than 20\% of cases, and Copilot Chat was unable to fix \textit{CWE-78: OS Command Injection}. In addition, we also observe that as the information provided in the prompts increases, the percentages of certain CWEs being fixed improve, such as \textit{CWE-79: Cross-site Scripting} and \textit{CWE-330: Use of Insufficiently Random Values Weakness}. The percentage of successful fixes for \textit{CWE-79: Cross-site Scripting} increases from 0\% to 70\%, while the percentage of successful fixes for \textit{CWE-330: Use of Insufficiently Random Values Weakness} increases from less than 10\% to 98\%.}

\section{Discussion}\label{Discussion}

\subsection{Interpretation of Results}
\label{interpretationofresults}
\noindent\textbf{RQ1: How secure is the code generated by Copilot in GitHub projects?} \\
\textcolor{black}{Around quarter of the code snippets in our dataset contained security weaknesses. We found that the code with security weaknesses in Python primarily involves Utility Tools and Game applications, whereas, in JavaScript, Web Application code constitutes the majority of security weaknesses followed by Utility Tools. This disparity may arise from the two languages' differing purposes and design philosophies~\cite{bootdev2024python,halfnine2024javascript}. JavaScript is mainly used for front-end development and Web applications~\cite{pubnub_javascript}, leading to its security weaknesses being concentrated in scenarios related to the web and user interaction. In contrast, Python is more broadly applied in fields such as data processing and scientific computing~\cite{nagpal2019python,simplilearn_python}, resulting in security weaknesses occurring across a wider range of application domains.}
Although our results show that the proportion of security weaknesses in Python code snippets is higher than in JavaScript, there is no significant difference between the two languages. Using AI code generation tools to write code in Python or JavaScript can generally lead to security weaknesses. This could be attributed to features that made their code more flexible, such as dynamic typing and interpretation. From a tool point of view, the generated code does not need to reason about the whole program because it is dynamically typed, and consequently, it does not need to read the flow of the whole program to generate working suggestions. In summary, developers should pay careful attention to the potential security weaknesses present in automatically generated code, taking appropriate measures to validate input data and manage resources effectively to minimize security risks. RQ1 results suggest that using Copilot is prone to security issues in production and additional security assessments are required to ensure the generated code does not introduce potential security risks.\\

\noindent\textbf{RQ2: What security weaknesses are present in the code generated by Copilot across different application domains?}\\
After conducting a security analysis of 733 code snippets generated by Copilot and other tools, a total of 628 security weaknesses were identified, involving 43 CWEs, which is around 10\% of the CWEs (439 CWEs)~\cite{cwe-software-development}. This may be because Copilot and the other two tools generate code in different programming languages and application scenarios. In addition, since the base models of AI code generation tools (such as Codex) are trained on publicly available data that potentially contain various types of security weaknesses, this can lead to the presence of multiple CWEs in the generated code. This set of 43 CWEs covers many security weaknesses. 

\textcolor{black}{The diversity of security weaknesses indicates that developers using Copilot face various security risks. These risks are diverse, covering different development environments and scenarios. At the same time, eight of the CWEs identified in our dataset can be found in the 2023 CWE Top-25 list, covering more than 200 security weaknesses (37.1\% of 628 identified). This indicates that the commonly acknowledged  Top-25 CWEs, which are considered the most prevalent and dangerous weaknesses, are also prevalent in the AI-generated code. Therefore, developers using Copilot must pay close attention to these weaknesses and take appropriate measures to prevent them before they are integrated into their code base. Meanwhile, we observed that some vulnerabilities from the CWE Top-25 list were not detected in our analysis, indicating that Copilot may sanitize and prevent specific weaknesses from being suggested to developers.} 

\textcolor{black}{We also identified 30 security weaknesses in the generated code that do not belong to the CWE Top-25 list. Although these are less common security weaknesses and may not be as widespread as CWE Top-25, attackers can still exploit them. For example, we only detected one instance of \textit{CWE-732: Incorrect Permission Assignment for Critical Resource} in our dataset. This security weakness is not commonly found in code and only occurs when specific users have certain permissions. However, it can lead to significant security risks when it does occur. }

\textcolor{black}{Besides, we observed that the types of CWEs in generated code are closely related to the application domains. Code snippets used in Game development are prone to issues such as \textit{CWE-330: Use of Insufficiently Random Values Weakness} and \textit{CWE-284: Improper Access Control}. This may be due to the complex logic and player inputs typically involved in game applications, which can lead to security problems related to memory management and input validation. In the Utility Tool category, the proportion of security weaknesses is relatively high, which may be attributed to the diversity of user inputs. Furthermore, we found that Python code snippets often exhibit security weaknesses such as \textit{CWE-330: Use of Insufficiently Random Values Weakness}, \textit{CWE-427: Uncontrolled Search Path Element}, and \textit{CWE-78: OS Command Injection}, whereas JavaScript code snippets primarily encounter issues like \textit{CWE-22: Path Traversal} and \textit{CWE-94: Code Injection}. This may be related to the differences in the application characteristics of the two languages: JavaScript, particularly through environments like Node.js, is use for file operations and dynamic code execution~\cite{chhetri2016comparative}, while Python tends to focus more on system-level operations and automation scripts. In addition, the measures we need to take are different: Utility Tools in JavaScript need to emphasize path validation and the security of dynamic execution to prevent attacks such as path traversal and code injection, while those in Python should focus on ensuring safe library loading paths and validating user input for command execution to prevent system-level vulnerabilities caused by malicious input. In Web development, Python is commonly used for frequent interactions between the back-end and the database~\cite{simplilearn_python}, which makes it prone to \textit{CWE-89: SQL Injection}. \textit{CWE-89: SQL Injection} typically occurs when user input is directly embedded into SQL query statements without parameterized queries or input validation. On the other hand, JavaScript, as a front-end language that interacts directly with users, often faces security weaknesses such as \textit{CWE-79: Cross-site Scripting}, \textit{CWE-94: Code Injection}, and \textit{CWE-22: Path Traversal}. This may be attributed to the frequent interactions between users and data in Web development. Additionally, the complexity of dependencies and the handling of sensitive data further increase the likelihood of security weaknesses.}

\textcolor{black}{Regarding the Network Communication category, the proportion of security weaknesses is relatively lower than other categories. This may be attributed to the fact that code in the Network Communication domain typically demands a high level of security, benefiting from mature protocols, stringent industry standards. Additionally, the protocols involved in network communication inherently incorporate certain security mechanisms to safeguard the data transmission process, further mitigating common attack risks.}

Overall, while Python and JavaScript differ in some common types of security weaknesses, they require developers to be aware of and take timely and targeted security measures to mitigate these risks. 
RQ2 findings note the security weaknesses that developers may encounter in an actual production environment and their occurrence frequency. These findings can help developers be aware of the security aspects of the code generated by AI code generation tools and take appropriate measures to address the security weaknesses in an informed manner.\\


\noindent\textbf{RQ3: Can \textit{Copilot Chat} help fix the security weaknesses found in the code generated by Copilot?} \\
\textcolor{black}{As shown in Table \ref{table:fix}, we can see that regardless of the prompts used, \textit{Copilot Chat} can fix a certain proportion of security weaknesses in the code we analyzed. Even the slash command \texttt{/fix} can resolve nearly 20\% of security problems, indicating that \textit{Copilot Chat} has some capability to fix certain weaknesses. Copilot Chat is an enhanced feature that extends the capabilities of the original Copilot~\cite{github_copilot_chat}. In fact, the security of Copilot and its underlying model (Codex) is continually improving~\cite{copilotxpreview}. However, it has also been found that Copilot's security layer can potentially handle some CWEs better than others, leaving applications vulnerable to some critical vulnerabilities~\cite{majdinasab2024assessing}. It is believed that similar programming-based chatbots, such as Amazon Q (which works with CodeWhisperer), also offer similar repair capabilities. Furthermore, we can observe that using prompts with richer information helps improve the repair effectiveness, which guides us to leverage static analysis tools in combination with \textit{Copilot Chat} to enhance the security of automatically generated code. When using the \/fix command, security weaknesses in JavaScript code are resolved more frequently than those in Python, possibly because JavaScript code is easier to trace in context, allowing \textit{Copilot Chat} to more readily identify and fix these issues. In contrast, Python often involves data structures, libraries, and system calls, which may make security weakness resolution less intuitive compared to JavaScript. However, when enhanced information is provided, the repair rates for security weaknesses in Python and JavaScript code segments do not differ significantly.}

\textcolor{black}{From the heatmap in Figure~\ref{fig:cwe-fix}, we can see that the effectiveness of \textit{Copilot Chat} in fixing different CWEs varies. \textit{Copilot Chat} is particularly effective at fixing \textit{CWE-259: Use of Hard-coded Password} and \textit{CWE-685: Function Call with Incorrect Number of Arguments}, which pertain to access control issues and environment configuration. These issues are relatively straightforward and have standard repair methods, making them easier for automated tools to identify and fix. On the other hand, other weaknesses such as \textit{CWE-78: command injection} and \textit{CWE-94: code injection} involve executing commands or code provided by users, which typically require better complex contextual understanding and detailed input validation. Due to the complexity of these issues, \textit{Copilot Chat} may encounter difficulties in automatically providing a fix. Additionally, we found that using more detailed instructions significantly improves repair outcomes for certain CWEs, such as \textit{CWE-330: Use of Insufficiently Random Values Weakness} and \textit{CWE-79: Cross-site Scripting}. This may be because our designed prompts provide richer information and context (the warning messages from static analysis tools), reducing ambiguity. This allows Copilot to effectively identify issues and generate correct fixes.}

\subsection{Implications}\label{implications}
\textcolor{black}{\subsubsection{Importance of continuous security analysis of automatically-generated code} We conjecture that practitioners using Copilot and other AI code generation tools are likely to encounter security weaknesses in their generated code, regardless of the programming language used. Our study results reflect the inevitability of security weaknesses in automatically generated code. Practitioners must be aware of the diverse security scenarios in production and adopt multiple security prevention measures to address security risks before accepting vulnerable code suggestions.} 

\textcolor{black}{Some CWEs, such as \textit{CWE-78: OS Command Injection} and \textit{CWE-94: Code Injection}, involve direct control of code execution by user inputs and may lead to severe security risks. Developers should be particularly vigilant about these high-risk operations and manually review these parts of the code. Input validation is the foundation of code security. Many security weaknesses (such as \textit{CWE-22: Path Traversal}) arise from inadequate input validation. When using AI code generation tools, developers should avoid directly using user input as the basis for code execution or file path operations and manually add input validation or adopt a more secure coding approach. Moreover, Copilot performs relatively well in handling permission control (such as \textit{CWE-259: Use of Hard-coded Password}) and environment configuration issues. Nevertheless, developers should be familiar with best practices related to permissions and verify that the configurations generated by AI code generation tools meet the project's security requirements.}

Maintaining a rigorous process of security analysis concurrently with code generation can help identify potential vulnerability issues and rectify the security weaknesses in time. Developers should follow the best practices for using code generation tools and always check the code suggestions generated by Copilot (or any code generation tools). For example, developers can establish a gated check-in build process to check and prevent security weaknesses when committing code generated by AI code generation tools. Initially, we can turn to automated tools to continuously scan the Copilot-generated code for known security weaknesses, such as CWEs. Then, \textit{Copilot Chat} or other LLMs can be used to fix identified code with security weaknesses, followed by a code re-analysis with security analysis tools. At the same time, the severity of security weaknesses and the limitations of static analysis tools remind us that it is essential to conduct a subsequent manual assessment, including manual security code review for auto-generated code. By embracing this combined strategy for continuous security analysis, we can ensure a robust security shield for the code-committing process with AI-generated code.

\textcolor{black}{\subsubsection{Prevention of security weaknesses in AI-generated code} We offer the following suggestions on how to prevent potential security weaknesses in generated code:}

\textcolor{black}{\textit{Targeted security countermeasures}: Based on the frequency of related CWEs in our dataset, practitioners can proactively prevent and address security weaknesses in a targeted manner. Developers should focus on dedicated CWEs in AI-generated code of different programming languages when using Copilot to generate e.g., Python or JavaScript code. Furthermore, a recent study shows that security weaknesses appear in certain code suggestions, but not all~\cite{majdinasab2024assessing}; consequently, developers should carefully select potentially more secure suggestions, with the assistance of tools, that do not expose the code to vulnerabilities. Security weaknesses can vary across programming languages and application domains, and developers should pay special attention to these differences when using these automated code generation tools. For example, the security weaknesses path validation (such as \textit{CWE-22: Path Traversal}) and dynamic execution (such as \textit{CWE-94: Code Injection}) are more likely to occur in JavaScript code. At the same time, system calls and library loading in Python code are more likely to raise security concerns.}

\textcolor{black}{\textit{Standardized security assessment}: Common security weaknesses in software development are also prevalent in the code generated by AI code generation tools. As a good practice, developers can use the CWE Top-25 list as a guide to understand which security weaknesses are most common and dangerous in the generated code. Additionally, the CWE Top-25 provides a standardized approach for security assessment, and developers can also use it as a guide to perform security audits of the AI-generated code. Using a standardized remediation strategy for common security weaknesses (such as privilege control and environment configuration) can effectively reduce the likelihood of generating insecure code. Developers can refer to the fixes provided in the static analysis tool documentation~\cite{codeqlhelp} or the mitigation measures for related CWEs provided by the MITRE~\cite{cwe}.}

\textcolor{black}{\textit{Enhancing prompt engineering to generate more secure code}: The instruction tuning schemes in code generation not only impact the utility of the code but also its security. By combining the security fine-tuning with standard instruction tuning, joint optimization of security and utility can be promoted~\cite{he2024instruction}. It is recommended to incorporate security considerations from the initial stages of code generation. For specific security scenarios prone to CWEs, we can improve code security by enhancing prompt engineering, e.g., with the prompt patterns proposed in~\cite{white2023prompt}. }

\textcolor{black}{\textit{Distinguishing the complexity of security weaknesses}: Some security issues, such as access control and environment configuration, have standard solutions and are well suited for quick fixes using tools like \textit{Copilot Chat}. However, for complex logic controls (such as logic processing related to multi-layered inputs), \textit{Copilot Chat}'s repair capabilities may be limited, requiring developers’ review and design. When using \textit{Copilot Chat}, developers should assess the complexity of issues to determine which code can be automatically fixed and which requires manual intervention.}

\textcolor{black}{\textit{Static analysis tools combined with LLM-based chatbots for fixing security weaknesses}: When using \textit{Copilot Chat} to address security weaknesses, developers should provide more specific context and clearer instructions. More detailed information can significantly enhance \textit{Copilot Chat}’s ability to fix security problems. Meanwhile, integrating static analysis tools allows for a more comprehensive identification of potential risks in the code. Developers can leverage static analysis tools to identify specific security risks and provide detailed instructions to Copilot Chat, enabling it to effectively fix the issues in Copilot-generated code.}

\section{Threats to Validity}\label{Threats to Validity}
The validity threats are discussed according to the guidelines in \cite{runeson2009guidelines}. Note that we did not consider internal validity threats since we did not investigate any relationships between variables and results.

\textbf{Construct Validity:} This study has three threats to construct validity: \textit{(1) Using keyword-based search --} We used a keyword-based search to collect relevant code snippets from GitHub. The results obtained through the keyword-based search may not cover all the code snippets generated by Copilot and other tools on GitHub. We tried to mitigate this threat by constantly and iteratively refining the keywords and using synonyms. \textcolor{black}{\textit{(2) Manual data filtering --} We manually screened the results obtained from the keyword-based search by analyzing the comments, tags, and other metadata of the code snippets to determine whether they were generated by Copilot or other tools. Since this process was manually done, it may have been influenced by personal bias. In this regard, two authors conducted the experiment independently to minimize the impact on the construct validity.} \textit{(3) Manual association of CWEs --} We manually associated the warning messages reported by the static analysis tools with a particular CWE, which may introduce personal subjective bias, threatening the construct validity. We employed two measures to mitigate this threat. First, since the list of CWEs is a tree structure with interconnections between them, we first matched the warning messages to a higher-level CWE and further checked whether we can match the warning messages to a lower-level CWE with a more specific definition. We refer to the description of the test suite and CWE coverage in the static analysis tool documentation. Second, to mitigate the personal bias, two authors independently assigned each security weakness description a CWE ID within a period of ten days. In case of disagreement, the two authors discussed it with the assessment by a third author (a security expert).

\textbf{External Validity:} \textcolor{black}{Our dataset consists of the code snippets generated by Copilot and other code generation tools collected from open-source projects on GitHub. During the filtering process, we excluded code that utilized AI code generation tools to solve simple programming practice problems, aiming to ensure that the collected data genuinely reflected open source development on GitHub. Since the data from GitHub are not diversified enough, we had a higher number of code snippets originating from the Game projects. This could result in a lack of comprehensiveness in the security scenarios involved. The peculiarity of the data source may make the dataset incomplete, thereby threatening the external validity of the results.
Furthermore, we acknowledge the need to collect more diverse code snippets from different platforms to increase the generalizability of the results. We will consider adopting more diversified ways or platforms to collect code.Additionally, due to the limitations of static analysis tools themselves, these tools could not scan all CWEs, and there is a degree of false positives in the analysis results (as the case with static analysis, in general, \cite{sui2020recall,kang2022detecting}). Although we used two widely used static analysis tools to increase the comprehensiveness of the scans and manually checked the results of the tool scans, the results may suffer from incompleteness.}

\textbf{Reliability:} We used multiple automated static analysis tools to analyze the Copilot-generated code snippets to improve security weaknesses detection. Developers have widely used these automated tools. The querying mechanism of these tools ensures that the analysis results remain consistent when used multiple times. In addition, we performed two rounds of analyzing with two tools for security checks on each code snippet, intending to complement the results of one tool with the other. By implementing these measures, we believe that our research results are reliable and these threats to reliability are mitigated.

\section{Conclusions}\label{Conclusions and Future Work}
Automatic code generation and recommendation have been an active research area due to the advancement of AI and, specifically, LLMs. AI code generation tools, such as Copilot, can significantly improve developers' development efficiency but can also introduce vulnerabilities and security risks. \textcolor{black}{Existing studies on the security of AI-generated code mainly focused on the security issues in the generated code using crafted scenarios, and potential security weaknesses of AI-generated code in practical scenarios in open source development environment (e.g., GitHub) have not been fully considered.} In this paper, we present the results of an empirical study to analyze security weaknesses in Copilot-generated code found in public GitHub projects. We identified 733 code snippets generated by Copilot and other tools (i.e., CodeWhisperer and Codeium) from GitHub projects and analyzed those snippets for security weaknesses using static analysis tools. We also examined using \textit{Copilot Chat} to fix security issues in Copilot-generated code. This study aims to help developers understand the security risks of weaknesses introduced in the code generated by Copilot (and potentially similar code generation tools). 
\textcolor{black}{Our results show that (1) around 30\% of the 733 generated code snippets contain security weaknesses. Developers have a high risk of raising security weaknesses when using Copilot or other AI code generation tools, regardless of the programming language, and therefore, applying appropriate security checks is necessary. (2) The detected security weaknesses are diverse in nature and are associated with 43 different CWEs. Developers encounter a variety of development scenarios and production environments, requiring appropriate security awareness and skills. (3) Among these CWEs, eight appear in the 2023 CWE Top-25 list, and six belong to the Stubborn Weaknesses, indicating high severity. (4) \textit{Copilot Chat} can help fix many of the security issues in Copilot-generated code, but its success rate varies across different CWEs. (5) Providing \textit{Copilot Chat} with a warning message from the static analysis tool leads to a better fix.}

In the future, we plan to: (1) collect additional code snippets from other open source repositories and industrial projects and code snippets generated by newer releases of Copilot; 
(2) analyze and summarize the application scenarios of these code snippets, studying how practitioners use Copilot and fix the issues in development; and (3) compare the results with other emerging Generative AI code generation tools such as CodeWhisperer, aiXcoder, and Code Llama.


\section*{Acknowledgments}
This work is funded by the National Natural Science Foundation of China (NSFC) under Grant No. 62172311 and the Major Science and Technology Project of Hubei Province under Grant No. 2024BAA008.

\end{sloppypar}

\bibliographystyle{ACM-Reference-Format}
\bibliography{ref}


\begin{thebibliography}{84}


\ifx \showCODEN    \undefined \def \showCODEN     #1{\unskip}     \fi
\ifx \showDOI      \undefined \def \showDOI       #1{#1}\fi
\ifx \showISBNx    \undefined \def \showISBNx     #1{\unskip}     \fi
\ifx \showISBNxiii \undefined \def \showISBNxiii  #1{\unskip}     \fi
\ifx \showISSN     \undefined \def \showISSN      #1{\unskip}     \fi
\ifx \showLCCN     \undefined \def \showLCCN      #1{\unskip}     \fi
\ifx \shownote     \undefined \def \shownote      #1{#1}          \fi
\ifx \showarticletitle \undefined \def \showarticletitle #1{#1}   \fi
\ifx \showURL      \undefined \def \showURL       {\relax}        \fi
\providecommand\bibfield[2]{#2}
\providecommand\bibinfo[2]{#2}
\providecommand\natexlab[1]{#1}
\providecommand\showeprint[2][]{arXiv:#2}

\bibitem[Asare et~al\mbox{.}(2023)]%
        {asare_is_2023}
\bibfield{author}{\bibinfo{person}{Owura Asare}, \bibinfo{person}{Meiyappan Nagappan}, {and} \bibinfo{person}{N. Asokan}.} \bibinfo{year}{2023}\natexlab{}.
\newblock \showarticletitle{Is GitHub's Copilot as Bad as Humans at Introducing Vulnerabilities in Code?}
\newblock \bibinfo{journal}{\emph{Empirical Software Engineering}} \bibinfo{volume}{28}, \bibinfo{number}{6} (\bibinfo{year}{2023}), \bibinfo{pages}{Article No. 129}.
\newblock
\urldef\tempurl%
\url{https://doi.org/10.1007/s10664-023-10380-1}
\showDOI{\tempurl}


\bibitem[Barke et~al\mbox{.}(2023)]%
        {barke2022grounded}
\bibfield{author}{\bibinfo{person}{Shraddha Barke}, \bibinfo{person}{Michael~B James}, {and} \bibinfo{person}{Nadia Polikarpova}.} \bibinfo{year}{2023}\natexlab{}.
\newblock \showarticletitle{Grounded copilot: How programmers interact with code-generating models}.
\newblock \bibinfo{journal}{\emph{Proceedings of the ACM on Programming Languages}} \bibinfo{volume}{7}, \bibinfo{number}{OOPSLA1} (\bibinfo{year}{2023}), \bibinfo{pages}{85--111}.
\newblock
\urldef\tempurl%
\url{https://doi.org/10.1145/3586030}
\showDOI{\tempurl}


\bibitem[Becker et~al\mbox{.}(2023)]%
        {becker2022programming}
\bibfield{author}{\bibinfo{person}{Brett~A Becker}, \bibinfo{person}{Paul Denny}, \bibinfo{person}{James Finnie-Ansley}, \bibinfo{person}{Andrew Luxton-Reilly}, \bibinfo{person}{James Prather}, {and} \bibinfo{person}{Eddie~Antonio Santos}.} \bibinfo{year}{2023}\natexlab{}.
\newblock \showarticletitle{Programming is hard-or at least it used to be: Educational opportunities and challenges of ai code generation}. In \bibinfo{booktitle}{\emph{Proceedings of the 54th ACM Technical Symposium on Computer Science Education V. 1}}. \bibinfo{pages}{500--506}.
\newblock
\urldef\tempurl%
\url{https://doi.org/10.1145/3545945.3569759}
\showDOI{\tempurl}


\bibitem[Brown et~al\mbox{.}(2020)]%
        {brown2020language}
\bibfield{author}{\bibinfo{person}{Tom Brown}, \bibinfo{person}{Benjamin Mann}, \bibinfo{person}{Nick Ryder}, \bibinfo{person}{Melanie Subbiah}, \bibinfo{person}{Jared~D Kaplan}, \bibinfo{person}{Prafulla Dhariwal}, \bibinfo{person}{Arvind Neelakantan}, \bibinfo{person}{Pranav Shyam}, \bibinfo{person}{Girish Sastry}, \bibinfo{person}{Amanda Askell}, {et~al\mbox{.}}} \bibinfo{year}{2020}\natexlab{}.
\newblock \showarticletitle{Language models are few-shot learners}. In \bibinfo{booktitle}{\emph{Proceedings of the 34th Annual Conference on Neural Information Processing Systems (NeurIPS)}}. \bibinfo{pages}{1877--1901}.
\newblock


\bibitem[Campbell et~al\mbox{.}(2013)]%
        {campbell2013coding}
\bibfield{author}{\bibinfo{person}{John~L Campbell}, \bibinfo{person}{Charles Quincy}, \bibinfo{person}{Jordan Osserman}, {and} \bibinfo{person}{Ove~K Pedersen}.} \bibinfo{year}{2013}\natexlab{}.
\newblock \showarticletitle{Coding in-depth semistructured interviews: Problems of unitization and intercoder reliability and agreement}.
\newblock \bibinfo{journal}{\emph{Sociological Methods \& Research}} \bibinfo{volume}{42}, \bibinfo{number}{3} (\bibinfo{year}{2013}), \bibinfo{pages}{294--320}.
\newblock
\urldef\tempurl%
\url{https://doi.org/10.1177/0049124113500475}
\showDOI{\tempurl}


\bibitem[Chen et~al\mbox{.}(2021)]%
        {chen2021evaluating}
\bibfield{author}{\bibinfo{person}{Mark Chen}, \bibinfo{person}{Jerry Tworek}, \bibinfo{person}{Heewoo Jun}, \bibinfo{person}{Qiming Yuan}, \bibinfo{person}{Henrique Ponde de~Oliveira Pinto}, \bibinfo{person}{Jared Kaplan}, \bibinfo{person}{Harri Edwards}, \bibinfo{person}{Yuri Burda}, \bibinfo{person}{Nicholas Joseph}, \bibinfo{person}{Greg Brockman}, {et~al\mbox{.}}} \bibinfo{year}{2021}\natexlab{}.
\newblock \showarticletitle{Evaluating large language models trained on code}.
\newblock \bibinfo{journal}{\emph{arXiv preprint arXiv:2107.03374}} (\bibinfo{year}{2021}).
\newblock
\urldef\tempurl%
\url{https://doi.org/10.48550/arXiv.2107.03374}
\showDOI{\tempurl}


\bibitem[Chhetri(2016)]%
        {chhetri2016comparative}
\bibfield{author}{\bibinfo{person}{Nimesh Chhetri}.} \bibinfo{year}{2016}\natexlab{}.
\newblock \showarticletitle{A comparative analysis of node. js (server-side javascript)}.
\newblock \bibinfo{journal}{\emph{Culminating Projects in Computer Science and Information Technology}}  \bibinfo{volume}{5} (\bibinfo{year}{2016}), \bibinfo{pages}{1--69}.
\newblock


\bibitem[CodeQL(2024)]%
        {codeqlhelp}
\bibfield{author}{\bibinfo{person}{CodeQL}.} \bibinfo{year}{2024}\natexlab{}.
\newblock \bibinfo{booktitle}{\emph{CodeQL Query Help}}.
\newblock
\urldef\tempurl%
\url{https://codeql.github.com/codeql-query-help/}
\showURL{%
\tempurl}


\bibitem[Cohen(1960)]%
        {Jacob1960Coefficient}
\bibfield{author}{\bibinfo{person}{Jacob Cohen}.} \bibinfo{year}{1960}\natexlab{}.
\newblock \showarticletitle{A Coefficient of Agreement for Nominal Scales}.
\newblock \bibinfo{journal}{\emph{Educational and Psychological Measurement}} \bibinfo{volume}{20}, \bibinfo{number}{1} (\bibinfo{year}{1960}), \bibinfo{pages}{37--46}.
\newblock
\urldef\tempurl%
\url{https://doi.org/10.1177/001316446002000104}
\showDOI{\tempurl}


\bibitem[Copilot(2024)]%
        {copilot-usage}
\bibfield{author}{\bibinfo{person}{Copilot}.} \bibinfo{year}{2024}\natexlab{}.
\newblock \bibinfo{booktitle}{\emph{best-practices-for-using-github-copilot}}.
\newblock
\urldef\tempurl%
\url{https://docs.github.com/zh/copilot/using-github-copilot/best-practices-for-using-github-copilot}
\showURL{%
\tempurl}


\bibitem[Cosentino et~al\mbox{.}(2017)]%
        {cosentino2017systematic}
\bibfield{author}{\bibinfo{person}{Valerio Cosentino}, \bibinfo{person}{Javier L~C{\'a}novas Izquierdo}, {and} \bibinfo{person}{Jordi Cabot}.} \bibinfo{year}{2017}\natexlab{}.
\newblock \showarticletitle{A systematic mapping study of software development with GitHub}.
\newblock \bibinfo{journal}{\emph{IEEE Access}}  \bibinfo{volume}{5} (\bibinfo{year}{2017}), \bibinfo{pages}{7173--7192}.
\newblock
\urldef\tempurl%
\url{https://doi.org/10.1109/ACCESS.2017.2682323}
\showDOI{\tempurl}


\bibitem[Dakhel et~al\mbox{.}(2023)]%
        {dakhel2022github}
\bibfield{author}{\bibinfo{person}{Arghavan~Moradi Dakhel}, \bibinfo{person}{Vahid Majdinasab}, \bibinfo{person}{Amin Nikanjam}, \bibinfo{person}{Foutse Khomh}, \bibinfo{person}{Michel~C Desmarais}, {and} \bibinfo{person}{Zhen Ming~Jack Jiang}.} \bibinfo{year}{2023}\natexlab{}.
\newblock \showarticletitle{Github copilot ai pair programmer: Asset or liability?}
\newblock \bibinfo{journal}{\emph{Journal of Systems and Software}}  \bibinfo{volume}{203} (\bibinfo{year}{2023}), \bibinfo{pages}{111734}.
\newblock
\urldef\tempurl%
\url{https://doi.org/10.1016/j.jss.2023.111734}
\showDOI{\tempurl}


\bibitem[Distefano et~al\mbox{.}(2019)]%
        {distefano2019scaling}
\bibfield{author}{\bibinfo{person}{Dino Distefano}, \bibinfo{person}{Manuel F{\"a}hndrich}, \bibinfo{person}{Francesco Logozzo}, {and} \bibinfo{person}{Peter~W O'Hearn}.} \bibinfo{year}{2019}\natexlab{}.
\newblock \showarticletitle{Scaling static analyses at Facebook}.
\newblock \bibinfo{journal}{\emph{Commun. ACM}} \bibinfo{volume}{62}, \bibinfo{number}{8} (\bibinfo{year}{2019}), \bibinfo{pages}{62--70}.
\newblock
\urldef\tempurl%
\url{https://doi.org/10.1145/3338112}
\showDOI{\tempurl}


\bibitem[Do et~al\mbox{.}(2020)]%
        {do2020software}
\bibfield{author}{\bibinfo{person}{Lisa Nguyen~Quang Do}, \bibinfo{person}{James~R Wright}, {and} \bibinfo{person}{Karim Ali}.} \bibinfo{year}{2020}\natexlab{}.
\newblock \showarticletitle{Why do software developers use static analysis tools? a user-centered study of developer needs and motivations}.
\newblock \bibinfo{journal}{\emph{IEEE Transactions on Software Engineering}} \bibinfo{volume}{48}, \bibinfo{number}{3} (\bibinfo{year}{2020}), \bibinfo{pages}{835--847}.
\newblock
\urldef\tempurl%
\url{https://doi.org/10.1109/TSE.2020.3004525}
\showDOI{\tempurl}


\bibitem[Dow(2021)]%
        {mirai_copilot_security}
\bibfield{author}{\bibinfo{person}{Alex Dow}.} \bibinfo{year}{2021}\natexlab{}.
\newblock \bibinfo{booktitle}{\emph{GitHub's Copilot Security Issues}}.
\newblock
\urldef\tempurl%
\url{https://miraisecurity.com/blog/githubs-copilot-security-issues}
\showURL{%
\tempurl}


\bibitem[Dunlap et~al\mbox{.}(2023)]%
        {dunlap2023finding}
\bibfield{author}{\bibinfo{person}{Trevor Dunlap}, \bibinfo{person}{Seaver Thorn}, \bibinfo{person}{William Enck}, {and} \bibinfo{person}{Bradley Reaves}.} \bibinfo{year}{2023}\natexlab{}.
\newblock \showarticletitle{Finding Fixed Vulnerabilities with Off-the-Shelf Static Analysis}. In \bibinfo{booktitle}{\emph{Proceedings of the 8th IEEE European Symposium on Security and Privacy (EuroS\&P)}}. IEEE, \bibinfo{pages}{489--505}.
\newblock
\urldef\tempurl%
\url{https://doi.org/10.1109/EuroSP57164.2023.00036}
\showDOI{\tempurl}


\bibitem[Easterbrook et~al\mbox{.}(2008)]%
        {easterbrook2008selecting}
\bibfield{author}{\bibinfo{person}{Steve Easterbrook}, \bibinfo{person}{Janice Singer}, \bibinfo{person}{Margaret-Anne Storey}, {and} \bibinfo{person}{Daniela Damian}.} \bibinfo{year}{2008}\natexlab{}.
\newblock \showarticletitle{Selecting Empirical Methods for Software Engineering Research}.
\newblock \bibinfo{publisher}{Springer}, \bibinfo{pages}{285--311}.
\newblock
\urldef\tempurl%
\url{https://doi.org/10.1007/978-1-84800-044-5_11}
\showDOI{\tempurl}


\bibitem[Elgedawy et~al\mbox{.}(2024)]%
        {elgedawy2024ocassionally}
\bibfield{author}{\bibinfo{person}{Ran Elgedawy}, \bibinfo{person}{John Sadik}, \bibinfo{person}{Senjuti Dutta}, \bibinfo{person}{Anuj Gautam}, \bibinfo{person}{Konstantinos Georgiou}, \bibinfo{person}{Farzin Gholamrezae}, \bibinfo{person}{Fujiao Ji}, \bibinfo{person}{Kyungchan Lim}, \bibinfo{person}{Qian Liu}, {and} \bibinfo{person}{Scott Ruoti}.} \bibinfo{year}{2024}\natexlab{}.
\newblock \showarticletitle{Ocassionally Secure: A Comparative Analysis of Code Generation Assistants}.
\newblock \bibinfo{journal}{\emph{arXiv preprint arXiv:2402.00689}} (\bibinfo{year}{2024}).
\newblock
\urldef\tempurl%
\url{https://doi.org/10.48550/arXiv.2402.00689}
\showDOI{\tempurl}


\bibitem[Ernst(2003)]%
        {ernst2003static}
\bibfield{author}{\bibinfo{person}{Michael~D Ernst}.} \bibinfo{year}{2003}\natexlab{}.
\newblock \showarticletitle{Static and dynamic analysis: Synergy and duality}. In \bibinfo{booktitle}{\emph{Proceedings of the ICSE Workshop on Dynamic Analysis (WODA)}}. ACM, \bibinfo{pages}{24--27}.
\newblock


\bibitem[{ESLint Contributors}(2024)]%
        {eslint}
\bibfield{author}{\bibinfo{person}{{ESLint Contributors}}.} \bibinfo{year}{2024}\natexlab{}.
\newblock \bibinfo{booktitle}{\emph{ESLint - Pluggable JavaScript linter}}.
\newblock GitHub.
\newblock
\urldef\tempurl%
\url{https://eslint.org/}
\showURL{%
\tempurl}


\bibitem[Fu et~al\mbox{.}(2024)]%
        {dataset}
\bibfield{author}{\bibinfo{person}{Yujia Fu}, \bibinfo{person}{Peng Liang}, \bibinfo{person}{Amjed Tahir}, \bibinfo{person}{Zengyang Li}, \bibinfo{person}{Mojtaba Shahin}, \bibinfo{person}{Jiaxin Yu}, {and} \bibinfo{person}{Jinfu Chen}.} \bibinfo{year}{2024}\natexlab{}.
\newblock \bibinfo{booktitle}{\emph{{Dataset of the Paper ``Security Weaknesses of Copilot-Generated Code in GitHub Projects: An Empirical Study''}}}.
\newblock
\urldef\tempurl%
\url{https://doi.org/10.5281/zenodo.10802054}
\showURL{%
\tempurl}


\bibitem[{GitHub}(2023)]%
        {codeql}
\bibfield{author}{\bibinfo{person}{{GitHub}}.} \bibinfo{year}{2023}\natexlab{}.
\newblock \bibinfo{booktitle}{\emph{CodeQL} (\bibinfo{edition}{1.6} ed.)}.
\newblock GitHub.
\newblock
\urldef\tempurl%
\url{https://securitylab.github.com/tools/codeql}
\showURL{%
\tempurl}


\bibitem[GitHub(2023)]%
        {github_copilot_chat}
\bibfield{author}{\bibinfo{person}{GitHub}.} \bibinfo{year}{2023}\natexlab{}.
\newblock \bibinfo{booktitle}{\emph{GitHub Copilot Chat Beta Now Available for Every Organization}}.
\newblock
\urldef\tempurl%
\url{https://github.blog/news-insights/product-news/github-copilot-chat-beta-now-available-for-every-organization/}
\showURL{%
\tempurl}


\bibitem[{GitHub}(2023)]%
        {githubcopilot}
\bibfield{author}{\bibinfo{person}{{GitHub}}.} \bibinfo{year}{2023}\natexlab{}.
\newblock \bibinfo{booktitle}{\emph{GitHub Copilot for Individuals}}.
\newblock
\urldef\tempurl%
\url{https://docs.github.com/en/copilot/overview-of-github-copilot/about-github-copilot-for-individuals}
\showURL{%
\tempurl}


\bibitem[GitHub(2023)]%
        {copilotxpreview}
\bibfield{author}{\bibinfo{person}{GitHub}.} \bibinfo{year}{2023}\natexlab{}.
\newblock \bibinfo{booktitle}{\emph{GitHub CopilotX Preview}}.
\newblock
\urldef\tempurl%
\url{https://github.com/features/preview/copilot-x}
\showURL{%
\tempurl}


\bibitem[{GitHub}(2023)]%
        {codeqlcli}
\bibfield{author}{\bibinfo{person}{{GitHub}}.} \bibinfo{year}{2023}\natexlab{}.
\newblock \bibinfo{booktitle}{\emph{Using the CodeQL CLI}}.
\newblock GitHub.
\newblock
\urldef\tempurl%
\url{https://docs.github.com/zh/code-security/codeql-cli/using-the-codeql-cli/analyzing-databases-with-the-codeql-cli}
\showURL{%
\tempurl}


\bibitem[GitHub(2024a)]%
        {copilot_chat}
\bibfield{author}{\bibinfo{person}{GitHub}.} \bibinfo{year}{2024}\natexlab{a}.
\newblock \bibinfo{booktitle}{\emph{GitHub Copilot Chat - Visual Studio Marketplace}}.
\newblock
\urldef\tempurl%
\url{https://marketplace.visualstudio.com/items?itemName=GitHub.copilot-chat}
\showURL{%
\tempurl}


\bibitem[GitHub(2024b)]%
        {octoverse2024}
\bibfield{author}{\bibinfo{person}{GitHub}.} \bibinfo{year}{2024}\natexlab{b}.
\newblock \bibinfo{booktitle}{\emph{Octoverse 2024 Top Programming Languages}}.
\newblock
\urldef\tempurl%
\url{https://github.blog/news-insights/octoverse/octoverse-2024/#the-most-popular-programming-languages}
\showURL{%
\tempurl}


\bibitem[Halfnine(2024)]%
        {halfnine2024javascript}
\bibfield{author}{\bibinfo{person}{Halfnine}.} \bibinfo{year}{2024}\natexlab{}.
\newblock \bibinfo{booktitle}{\emph{JavaScript vs Python: Breaking Down Performance, Ease of Use, and More}}.
\newblock
\urldef\tempurl%
\url{https://www.halfnine.com/blog/post/javascript-vs-python}
\showURL{%
\tempurl}


\bibitem[Harding and Kloster(2024)]%
        {harding2024coding}
\bibfield{author}{\bibinfo{person}{William Harding} {and} \bibinfo{person}{Matthew Kloster}.} \bibinfo{year}{2024}\natexlab{}.
\newblock \bibinfo{booktitle}{\emph{Coding on Copilot: 2023 Data Suggests Downward Pressure on Code Quality}}.
\newblock
\urldef\tempurl%
\url{https://www.gitclear.com/coding_on_copilot_data_shows_ais_downward_pressure_on_code_quality}
\showURL{%
\tempurl}


\bibitem[He and Vechev(2023)]%
        {he2023controlling}
\bibfield{author}{\bibinfo{person}{Jingxuan He} {and} \bibinfo{person}{Martin Vechev}.} \bibinfo{year}{2023}\natexlab{}.
\newblock \showarticletitle{Controlling Large Language Models to Generate Secure and Vulnerable Code}.
\newblock \bibinfo{journal}{\emph{arXiv preprint arXiv:2302.05319}} (\bibinfo{year}{2023}).
\newblock
\urldef\tempurl%
\url{https://doi.org/10.48550/arXiv.2302.05319}
\showDOI{\tempurl}


\bibitem[He et~al\mbox{.}(2024)]%
        {he2024instruction}
\bibfield{author}{\bibinfo{person}{Jingxuan He}, \bibinfo{person}{Mark Vero}, \bibinfo{person}{Gabriela Krasnopolska}, {and} \bibinfo{person}{Martin Vechev}.} \bibinfo{year}{2024}\natexlab{}.
\newblock \showarticletitle{Instruction Tuning for Secure Code Generation}.
\newblock \bibinfo{journal}{\emph{arXiv preprint arXiv:2402.09497}} (\bibinfo{year}{2024}).
\newblock
\urldef\tempurl%
\url{https://doi.org/10.48550/arXiv.2402.09497}
\showDOI{\tempurl}


\bibitem[Huth and Nielson(2019)]%
        {huth2019static}
\bibfield{author}{\bibinfo{person}{Michael Huth} {and} \bibinfo{person}{Flemming Nielson}.} \bibinfo{year}{2019}\natexlab{}.
\newblock \bibinfo{title}{Static analysis for proactive security}.
\newblock , \bibinfo{numpages}{374--392}~pages.
\newblock
\urldef\tempurl%
\url{https://doi.org/10.1007/978-3-319-91908-9_19}
\showDOI{\tempurl}


\bibitem[Iannone et~al\mbox{.}(2022)]%
        {iannone2022secret}
\bibfield{author}{\bibinfo{person}{Emanuele Iannone}, \bibinfo{person}{Roberta Guadagni}, \bibinfo{person}{Filomena Ferrucci}, \bibinfo{person}{Andrea De~Lucia}, {and} \bibinfo{person}{Fabio Palomba}.} \bibinfo{year}{2022}\natexlab{}.
\newblock \showarticletitle{The secret life of software vulnerabilities: A large-scale empirical study}.
\newblock \bibinfo{journal}{\emph{IEEE Transactions on Software Engineering}} \bibinfo{volume}{49}, \bibinfo{number}{1} (\bibinfo{year}{2022}), \bibinfo{pages}{44--63}.
\newblock
\urldef\tempurl%
\url{https://doi.org/10.1109/TSE.2022.3140868}
\showDOI{\tempurl}


\bibitem[Kang et~al\mbox{.}(2022)]%
        {kang2022detecting}
\bibfield{author}{\bibinfo{person}{Hong~Jin Kang}, \bibinfo{person}{Khai~Loong Aw}, {and} \bibinfo{person}{David Lo}.} \bibinfo{year}{2022}\natexlab{}.
\newblock \showarticletitle{Detecting false alarms from automatic static analysis tools: How far are we?}. In \bibinfo{booktitle}{\emph{Proceedings of the 44th International Conference on Software Engineering (ICSE)}}. ACM, \bibinfo{pages}{698--709}.
\newblock
\urldef\tempurl%
\url{https://doi.org/10.1145/3510003.3510214}
\showDOI{\tempurl}


\bibitem[Kaur and Nayyar(2020)]%
        {kaur2020comparative}
\bibfield{author}{\bibinfo{person}{Arvinder Kaur} {and} \bibinfo{person}{Ruchikaa Nayyar}.} \bibinfo{year}{2020}\natexlab{}.
\newblock \showarticletitle{A comparative study of static code analysis tools for vulnerability detection in c/c++ and java source code}.
\newblock \bibinfo{journal}{\emph{Procedia Computer Science}}  \bibinfo{volume}{171} (\bibinfo{year}{2020}), \bibinfo{pages}{2023--2029}.
\newblock
\urldef\tempurl%
\url{https://doi.org/10.1016/j.procs.2020.04.217}
\showDOI{\tempurl}


\bibitem[Khoury et~al\mbox{.}(2023)]%
        {khoury2023secure}
\bibfield{author}{\bibinfo{person}{Rapha{\"e}l Khoury}, \bibinfo{person}{Anderson~R Avila}, \bibinfo{person}{Jacob Brunelle}, {and} \bibinfo{person}{Baba~Mamadou Camara}.} \bibinfo{year}{2023}\natexlab{}.
\newblock \showarticletitle{How secure is code generated by chatgpt?}. In \bibinfo{booktitle}{\emph{Proceedings of the IEEE International Conference on Systems, Man, and Cybernetics (SMC)}}. IEEE, \bibinfo{pages}{2445--2451}.
\newblock
\urldef\tempurl%
\url{https://doi.org/10.1109/SMC53992.2023.10394237}
\showDOI{\tempurl}


\bibitem[Krill(2024)]%
        {infoworld2024}
\bibfield{author}{\bibinfo{person}{Paul Krill}.} \bibinfo{year}{2024}\natexlab{}.
\newblock \bibinfo{booktitle}{\emph{{GitHub Copilot makes insecure code even less secure, Snyk says}}}.
\newblock
\urldef\tempurl%
\url{https://www.infoworld.com/article/3713141/github-copilot-makes-insecure-code-even-less-secure-snyk-says.amp.html}
\showURL{%
\tempurl}


\bibitem[Lertbanjongngam et~al\mbox{.}(2022)]%
        {lertbanjongngam2022empirical}
\bibfield{author}{\bibinfo{person}{Sila Lertbanjongngam}, \bibinfo{person}{Bodin Chinthanet}, \bibinfo{person}{Takashi Ishio}, \bibinfo{person}{Raula~Gaikovina Kula}, \bibinfo{person}{Pattara Leelaprute}, \bibinfo{person}{Bundit Manaskasemsak}, \bibinfo{person}{Arnon Rungsawang}, {and} \bibinfo{person}{Kenichi Matsumoto}.} \bibinfo{year}{2022}\natexlab{}.
\newblock \showarticletitle{An Empirical Evaluation of Competitive Programming AI: A Case Study of AlphaCode}. In \bibinfo{booktitle}{\emph{Proceedings of the 16th IEEE International Workshop on Software Clones (IWSC)}}. IEEE, \bibinfo{pages}{10--15}.
\newblock
\urldef\tempurl%
\url{https://doi.org/10.48550/arXiv.2208.08603}
\showDOI{\tempurl}


\bibitem[Li et~al\mbox{.}(2022)]%
        {li2022competition}
\bibfield{author}{\bibinfo{person}{Yujia Li}, \bibinfo{person}{David Choi}, \bibinfo{person}{Junyoung Chung}, \bibinfo{person}{Nate Kushman}, \bibinfo{person}{Julian Schrittwieser}, \bibinfo{person}{R{\'e}mi Leblond}, \bibinfo{person}{Tom Eccles}, \bibinfo{person}{James Keeling}, \bibinfo{person}{Felix Gimeno}, \bibinfo{person}{Agustin Dal~Lago}, {et~al\mbox{.}}} \bibinfo{year}{2022}\natexlab{}.
\newblock \showarticletitle{Competition-level code generation with alphacode}.
\newblock \bibinfo{journal}{\emph{Science}} \bibinfo{volume}{378}, \bibinfo{number}{6624} (\bibinfo{year}{2022}), \bibinfo{pages}{1092--1097}.
\newblock
\urldef\tempurl%
\url{https://doi.org/10.1126/science.abq1158}
\showDOI{\tempurl}


\bibitem[Li et~al\mbox{.}(2023)]%
        {li2022cctest}
\bibfield{author}{\bibinfo{person}{Zongjie Li}, \bibinfo{person}{Chaozheng Wang}, \bibinfo{person}{Zhibo Liu}, \bibinfo{person}{Haoxuan Wang}, \bibinfo{person}{Dong Chen}, \bibinfo{person}{Shuai Wang}, {and} \bibinfo{person}{Cuiyun Gao}.} \bibinfo{year}{2023}\natexlab{}.
\newblock \showarticletitle{Cctest: Testing and repairing code completion systems}. In \bibinfo{booktitle}{\emph{2023 IEEE/ACM 45th International Conference on Software Engineering (ICSE)}}. IEEE, \bibinfo{pages}{1238--1250}.
\newblock
\urldef\tempurl%
\url{https://doi.org/10.1109/ICSE48619.2023.00110}
\showDOI{\tempurl}


\bibitem[Liu et~al\mbox{.}(2024)]%
        {liu2023refining}
\bibfield{author}{\bibinfo{person}{Yue Liu}, \bibinfo{person}{Thanh Le-Cong}, \bibinfo{person}{Ratnadira Widyasari}, \bibinfo{person}{Chakkrit Tantithamthavorn}, \bibinfo{person}{Li Li}, \bibinfo{person}{Xuan-Bach~D Le}, {and} \bibinfo{person}{David Lo}.} \bibinfo{year}{2024}\natexlab{}.
\newblock \showarticletitle{Refining chatgpt-generated code: Characterizing and mitigating code quality issues}.
\newblock \bibinfo{journal}{\emph{ACM Transactions on Software Engineering and Methodology}} \bibinfo{volume}{33}, \bibinfo{number}{5} (\bibinfo{year}{2024}), \bibinfo{pages}{1--26}.
\newblock
\urldef\tempurl%
\url{https://doi.org/10.1145/3643674}
\showDOI{\tempurl}


\bibitem[Majdinasab et~al\mbox{.}(2024)]%
        {majdinasab2024assessing}
\bibfield{author}{\bibinfo{person}{Vahid Majdinasab}, \bibinfo{person}{Michael~Joshua Bishop}, \bibinfo{person}{Shawn Rasheed}, \bibinfo{person}{Arghavan Moradidakhel}, \bibinfo{person}{Amjed Tahir}, {and} \bibinfo{person}{Foutse Khomh}.} \bibinfo{year}{2024}\natexlab{}.
\newblock \showarticletitle{Assessing the Security of GitHub Copilot Generated Code - A Targeted Replication Study}. In \bibinfo{booktitle}{\emph{Proceedings of the 31st IEEE International Conference on Software Analysis, Evolution and Reengineering (SANER)}}. IEEE, \bibinfo{pages}{435--444}.
\newblock
\urldef\tempurl%
\url{https://doi.org/10.1109/SANER60148.2024.00051}
\showDOI{\tempurl}


\bibitem[McDaniel(2021)]%
        {gitguardian2021github}
\bibfield{author}{\bibinfo{person}{Dwayne McDaniel}.} \bibinfo{year}{2021}\natexlab{}.
\newblock \bibinfo{booktitle}{\emph{GitHub Copilot: Security and Privacy}}.
\newblock
\urldef\tempurl%
\url{https://blog.gitguardian.com/github-copilot-security-and-privacy/}
\showURL{%
\tempurl}


\bibitem[MITRE(2023)]%
        {cwe2023}
\bibfield{author}{\bibinfo{person}{MITRE}.} \bibinfo{year}{2023}\natexlab{}.
\newblock \bibinfo{booktitle}{\emph{CWE Top 25 Most Dangerous Software Errors 2023}}.
\newblock
\urldef\tempurl%
\url{https://cwe.mitre.org/top25/archive/2023/2023_top25_list.html}
\showURL{%
\tempurl}


\bibitem[MITRE(2024)]%
        {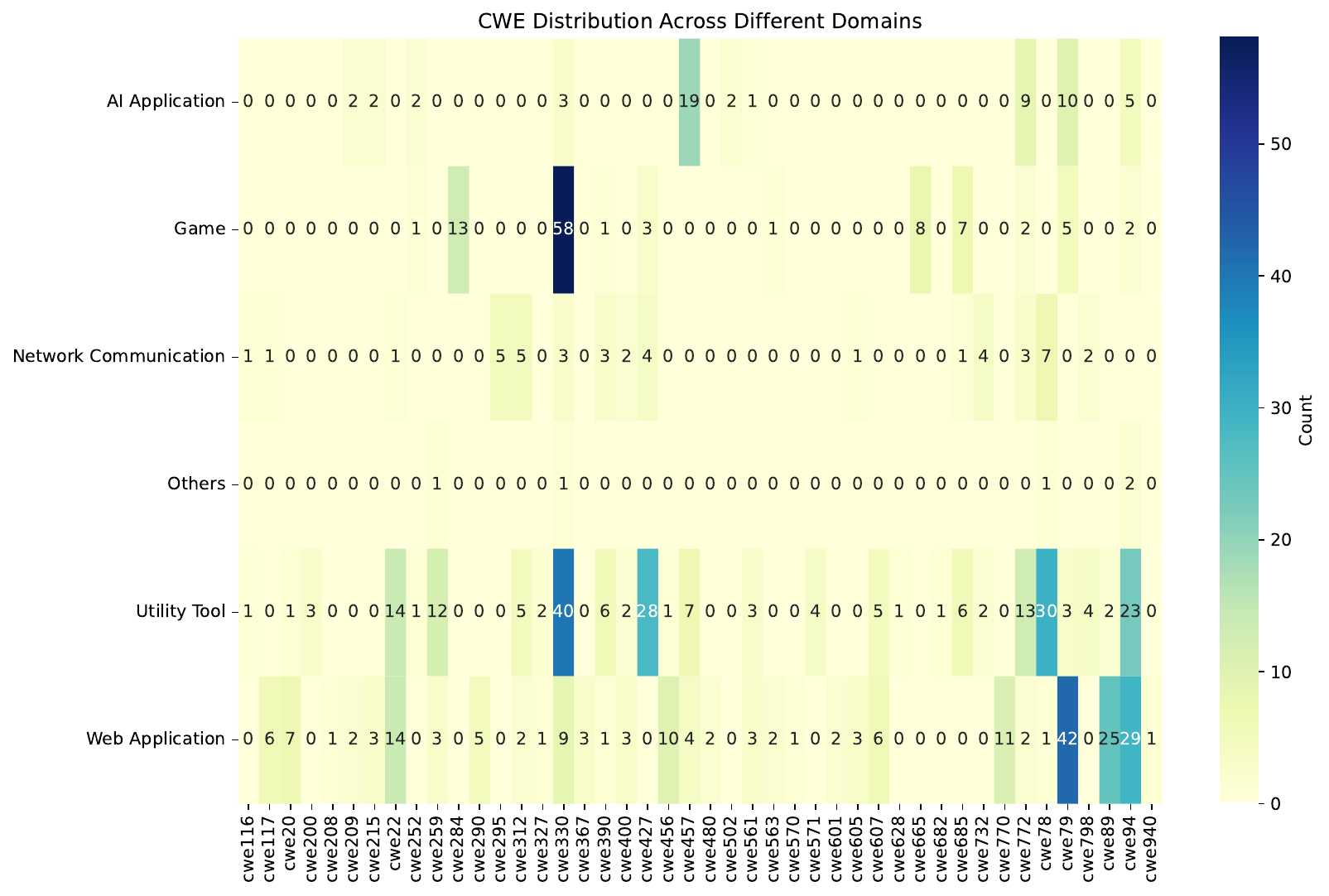}
\bibfield{author}{\bibinfo{person}{MITRE}.} \bibinfo{year}{2024}\natexlab{}.
\newblock \bibinfo{booktitle}{\emph{Common Weakness Enumeration (CWE) Data}}.
\newblock
\urldef\tempurl%
\url{https://cwe.mitre.org/data/index.html}
\showURL{%
\tempurl}


\bibitem[Mozannar et~al\mbox{.}(2024)]%
        {mozannar2022reading}
\bibfield{author}{\bibinfo{person}{Hussein Mozannar}, \bibinfo{person}{Gagan Bansal}, \bibinfo{person}{Adam Fourney}, {and} \bibinfo{person}{Eric Horvitz}.} \bibinfo{year}{2024}\natexlab{}.
\newblock \showarticletitle{Reading between the lines: Modeling user behavior and costs in AI-assisted programming}. In \bibinfo{booktitle}{\emph{Proceedings of the CHI Conference on Human Factors in Computing Systems}}. \bibinfo{pages}{1--16}.
\newblock
\urldef\tempurl%
\url{https://doi.org/10.1145/3613904.3641936}
\showDOI{\tempurl}


\bibitem[Nagpal and Gabrani(2019)]%
        {nagpal2019python}
\bibfield{author}{\bibinfo{person}{Abhinav Nagpal} {and} \bibinfo{person}{Goldie Gabrani}.} \bibinfo{year}{2019}\natexlab{}.
\newblock \showarticletitle{Python for data analytics, scientific and technical applications}. In \bibinfo{booktitle}{\emph{Proceedings of the Amity international conference on artificial intelligence (AICAI)}}. IEEE, \bibinfo{pages}{140--145}.
\newblock
\urldef\tempurl%
\url{https://doi.org/10.1109/AICAI.2019.8701341}
\showDOI{\tempurl}


\bibitem[Natella et~al\mbox{.}(2024)]%
        {natella2024ai}
\bibfield{author}{\bibinfo{person}{Roberto Natella}, \bibinfo{person}{Pietro Liguori}, \bibinfo{person}{Cristina Improta}, \bibinfo{person}{Bojan Cukic}, {and} \bibinfo{person}{Domenico Cotroneo}.} \bibinfo{year}{2024}\natexlab{}.
\newblock \showarticletitle{AI Code Generators for Security: Friend or Foe?}
\newblock \bibinfo{journal}{\emph{IEEE Security \& Privacy}} \bibinfo{volume}{22}, \bibinfo{number}{5} (\bibinfo{year}{2024}), \bibinfo{pages}{73--81}.
\newblock
\urldef\tempurl%
\url{https://doi.org/10.1109/MSEC.2024.3355713}
\showDOI{\tempurl}


\bibitem[Nguyen and Nadi(2022)]%
        {nguyen2022empirical}
\bibfield{author}{\bibinfo{person}{Nhan Nguyen} {and} \bibinfo{person}{Sarah Nadi}.} \bibinfo{year}{2022}\natexlab{}.
\newblock \showarticletitle{An empirical evaluation of GitHub copilot's code suggestions}. In \bibinfo{booktitle}{\emph{Proceedings of the 19th IEEE/ACM International Conference on Mining Software Repositories (MSR)}}. IEEE, \bibinfo{pages}{1--5}.
\newblock
\urldef\tempurl%
\url{https://doi.org/10.1145/3524842.3528470}
\showDOI{\tempurl}


\bibitem[Nunes et~al\mbox{.}(2018)]%
        {nunes2018comment}
\bibfield{author}{\bibinfo{person}{Joao~Pedro Nunes}, \bibinfo{person}{Alex~S Ribeiro}, \bibinfo{person}{Brad~J Schoenfeld}, {and} \bibinfo{person}{Edilson~S Cyrino}.} \bibinfo{year}{2018}\natexlab{}.
\newblock \showarticletitle{Comment on: “Comparison of periodized and non-periodized resistance training on maximal strength: A meta-analysis”}.
\newblock \bibinfo{journal}{\emph{Sports Medicine}}  \bibinfo{volume}{48} (\bibinfo{year}{2018}), \bibinfo{pages}{491--494}.
\newblock
\urldef\tempurl%
\url{https://doi.org/10.1007/s40279-017-0824-x}
\showDOI{\tempurl}


\bibitem[OpenAI(2021)]%
        {codex}
\bibfield{author}{\bibinfo{person}{OpenAI}.} \bibinfo{year}{2021}\natexlab{}.
\newblock \bibinfo{booktitle}{\emph{Introducing Codex: The AI Behind GitHub Copilot}}.
\newblock
\urldef\tempurl%
\url{https://openai.com/blog/openai-codex}
\showURL{%
\tempurl}


\bibitem[Overflow(2024)]%
        {stackoverflow2024survey}
\bibfield{author}{\bibinfo{person}{Stack Overflow}.} \bibinfo{year}{2024}\natexlab{}.
\newblock \bibinfo{booktitle}{\emph{Stack Overflow Developer Survey 2024: Most Popular Technologies (AI, Search, Dev)}}.
\newblock
\urldef\tempurl%
\url{https://survey.stackoverflow.co/2024/technology#most-popular-technologies-ai-search-dev}
\showURL{%
\tempurl}


\bibitem[{OWASP}(2024)]%
        {owasp}
\bibfield{author}{\bibinfo{person}{{OWASP}}.} \bibinfo{year}{2024}\natexlab{}.
\newblock \bibinfo{booktitle}{\emph{Source Code Analysis Tools}}.
\newblock
\urldef\tempurl%
\url{https://owasp.org/www-community/Source_Code_Analysis_Tools}
\showURL{%
\tempurl}


\bibitem[Pearce et~al\mbox{.}(2022)]%
        {pearce2022asleep}
\bibfield{author}{\bibinfo{person}{Hammond Pearce}, \bibinfo{person}{Baleegh Ahmad}, \bibinfo{person}{Benjamin Tan}, \bibinfo{person}{Brendan Dolan-Gavitt}, {and} \bibinfo{person}{Ramesh Karri}.} \bibinfo{year}{2022}\natexlab{}.
\newblock \showarticletitle{Asleep at the keyboard? assessing the security of github copilot’s code contributions}. In \bibinfo{booktitle}{\emph{Proceedings of the 43rd IEEE Symposium on Security and Privacy (SP)}}. IEEE, \bibinfo{pages}{754--768}.
\newblock
\urldef\tempurl%
\url{https://doi.org/10.1109/SP46214.2022.9833571}
\showDOI{\tempurl}


\bibitem[Perry et~al\mbox{.}(2023)]%
        {perry2022users}
\bibfield{author}{\bibinfo{person}{Neil Perry}, \bibinfo{person}{Megha Srivastava}, \bibinfo{person}{Deepak Kumar}, {and} \bibinfo{person}{Dan Boneh}.} \bibinfo{year}{2023}\natexlab{}.
\newblock \showarticletitle{Do users write more insecure code with AI assistants?}. In \bibinfo{booktitle}{\emph{Proceedings of the 2023 ACM SIGSAC Conference on Computer and Communications Security}}. \bibinfo{pages}{2785--2799}.
\newblock
\urldef\tempurl%
\url{https://doi.org/10.1145/3576915.3623157}
\showDOI{\tempurl}


\bibitem[PubNub(2024)]%
        {pubnub_javascript}
\bibfield{author}{\bibinfo{person}{PubNub}.} \bibinfo{year}{2024}\natexlab{}.
\newblock \bibinfo{booktitle}{\emph{JavaScript: The Complete Guide for Front-End Development}}.
\newblock
\urldef\tempurl%
\url{https://www.pubnub.com/guides/javascript/}
\showURL{%
\tempurl}


\bibitem[Pudari and Ernst(2023)]%
        {pudari2023copilot}
\bibfield{author}{\bibinfo{person}{Rohith Pudari} {and} \bibinfo{person}{Neil~A Ernst}.} \bibinfo{year}{2023}\natexlab{}.
\newblock \showarticletitle{From Copilot to Pilot: Towards AI Supported Software Development}.
\newblock \bibinfo{journal}{\emph{arXiv preprint arXiv:2303.04142}} (\bibinfo{year}{2023}).
\newblock
\urldef\tempurl%
\url{https://doi.org/10.48550/arXiv.2303.04142}
\showDOI{\tempurl}


\bibitem[PyCQA(2024)]%
        {bandit}
\bibfield{author}{\bibinfo{person}{PyCQA}.} \bibinfo{year}{2024}\natexlab{}.
\newblock \bibinfo{booktitle}{\emph{Bandit: A Security Linter for Python}}.
\newblock
\urldef\tempurl%
\url{https://github.com/PyCQA/bandit}
\showURL{%
\tempurl}


\bibitem[Reichenbach(2022)]%
        {bootdev2024python}
\bibfield{author}{\bibinfo{person}{Meghan Reichenbach}.} \bibinfo{year}{2022}\natexlab{}.
\newblock \bibinfo{booktitle}{\emph{Python vs. JavaScript: Which Should You Learn?}}
\newblock
\urldef\tempurl%
\url{https://blog.boot.dev/python/python-vs-javascript/}
\showURL{%
\tempurl}


\bibitem[Rokon et~al\mbox{.}(2020)]%
        {rokon2020sourcefinder}
\bibfield{author}{\bibinfo{person}{Md~Omar~Faruk Rokon}, \bibinfo{person}{Risul Islam}, \bibinfo{person}{Ahmad Darki}, \bibinfo{person}{Evangelos~E Papalexakis}, {and} \bibinfo{person}{Michalis Faloutsos}.} \bibinfo{year}{2020}\natexlab{}.
\newblock \showarticletitle{SourceFinder: Finding Malware Source-Code from Publicly Available Repositories in GitHub}. In \bibinfo{booktitle}{\emph{Proceedings of the 23rd International Symposium on Research in Attacks, Intrusions and Defenses (RAID)}}. USENIX, \bibinfo{pages}{149--163}.
\newblock


\bibitem[Runeson and H{\"o}st(2009)]%
        {runeson2009guidelines}
\bibfield{author}{\bibinfo{person}{Per Runeson} {and} \bibinfo{person}{Martin H{\"o}st}.} \bibinfo{year}{2009}\natexlab{}.
\newblock \showarticletitle{Guidelines for conducting and reporting case study research in software engineering}.
\newblock \bibinfo{journal}{\emph{Empirical Software Engineering}}  \bibinfo{volume}{14} (\bibinfo{year}{2009}), \bibinfo{pages}{131--164}.
\newblock
\urldef\tempurl%
\url{https://doi.org/10.1007/s10664-008-9102-8}
\showDOI{\tempurl}


\bibitem[Sadowski et~al\mbox{.}(2018)]%
        {sadowski2018lessons}
\bibfield{author}{\bibinfo{person}{Caitlin Sadowski}, \bibinfo{person}{Edward Aftandilian}, \bibinfo{person}{Alex Eagle}, \bibinfo{person}{Liam Miller-Cushon}, {and} \bibinfo{person}{Ciera Jaspan}.} \bibinfo{year}{2018}\natexlab{}.
\newblock \showarticletitle{Lessons from building static analysis tools at google}.
\newblock \bibinfo{journal}{\emph{Commun. ACM}} \bibinfo{volume}{61}, \bibinfo{number}{4} (\bibinfo{year}{2018}), \bibinfo{pages}{58--66}.
\newblock
\urldef\tempurl%
\url{https://doi.org/10.1145/3188720}
\showDOI{\tempurl}


\bibitem[Sandoval et~al\mbox{.}(2022)]%
        {sandoval2022security}
\bibfield{author}{\bibinfo{person}{Gustavo Sandoval}, \bibinfo{person}{Hammond Pearce}, \bibinfo{person}{Teo Nys}, \bibinfo{person}{Ramesh Karri}, \bibinfo{person}{Brendan Dolan-Gavitt}, {and} \bibinfo{person}{Siddharth Garg}.} \bibinfo{year}{2022}\natexlab{}.
\newblock \showarticletitle{Security Implications of Large Language Model Code Assistants: A User Study}.
\newblock \bibinfo{journal}{\emph{arXiv preprint arXiv:2208.09727}} (\bibinfo{year}{2022}).
\newblock
\urldef\tempurl%
\url{https://doi.org/10.48550/arXiv.2208.09727}
\showDOI{\tempurl}


\bibitem[Sarkar et~al\mbox{.}(2022)]%
        {sarkar2022like}
\bibfield{author}{\bibinfo{person}{Advait Sarkar}, \bibinfo{person}{Andrew~D Gordon}, \bibinfo{person}{Carina Negreanu}, \bibinfo{person}{Christian Poelitz}, \bibinfo{person}{Sruti~Srinivasa Ragavan}, {and} \bibinfo{person}{Ben Zorn}.} \bibinfo{year}{2022}\natexlab{}.
\newblock \showarticletitle{What is it like to program with artificial intelligence?}
\newblock \bibinfo{journal}{\emph{arXiv preprint arXiv:2208.06213}} (\bibinfo{year}{2022}).
\newblock
\urldef\tempurl%
\url{https://doi.org/10.48550/arXiv.2208.06213}
\showDOI{\tempurl}


\bibitem[She et~al\mbox{.}(2023)]%
        {she2023pitfalls}
\bibfield{author}{\bibinfo{person}{Xinyu She}, \bibinfo{person}{Yue Liu}, \bibinfo{person}{Yanjie Zhao}, \bibinfo{person}{Yiling He}, \bibinfo{person}{Li Li}, \bibinfo{person}{Chakkrit Tantithamthavorn}, \bibinfo{person}{Zhan Qin}, {and} \bibinfo{person}{Haoyu Wang}.} \bibinfo{year}{2023}\natexlab{}.
\newblock \showarticletitle{Pitfalls in Language Models for Code Intelligence: A Taxonomy and Survey}.
\newblock \bibinfo{journal}{\emph{arXiv preprint arXiv:2310.17903}} (\bibinfo{year}{2023}).
\newblock
\urldef\tempurl%
\url{https://doi.org/10.48550/arXiv.2310.17903}
\showDOI{\tempurl}


\bibitem[Siddiq et~al\mbox{.}(2022)]%
        {siddiq2022empirical}
\bibfield{author}{\bibinfo{person}{Mohammed~Latif Siddiq}, \bibinfo{person}{Shafayat~H Majumder}, \bibinfo{person}{Maisha~R Mim}, \bibinfo{person}{Sourov Jajodia}, {and} \bibinfo{person}{Joanna~CS Santos}.} \bibinfo{year}{2022}\natexlab{}.
\newblock \showarticletitle{An Empirical Study of Code Smells in Transformer-based Code Generation Techniques}. In \bibinfo{booktitle}{\emph{Proceedings of the 22nd IEEE International Working Conference on Source Code Analysis and Manipulation (SCAM)}}. IEEE, \bibinfo{pages}{71--82}.
\newblock
\urldef\tempurl%
\url{https://doi.org/10.1109/SCAM55253.2022.00014}
\showDOI{\tempurl}


\bibitem[Siddiq and Santos(2022)]%
        {siddiq2022securityeval}
\bibfield{author}{\bibinfo{person}{Mohammed~Latif Siddiq} {and} \bibinfo{person}{Joanna~CS Santos}.} \bibinfo{year}{2022}\natexlab{}.
\newblock \showarticletitle{SecurityEval dataset: mining vulnerability examples to evaluate machine learning-based code generation techniques}. In \bibinfo{booktitle}{\emph{Proceedings of the 1st International Workshop on Mining Software Repositories Applications for Privacy and Security (MSR4P\&S)}}. ACM, \bibinfo{pages}{29--33}.
\newblock
\urldef\tempurl%
\url{https://doi.org/10.1145/3549035.3561184}
\showDOI{\tempurl}


\bibitem[Simplilearn(2024)]%
        {simplilearn_python}
\bibfield{author}{\bibinfo{person}{Simplilearn}.} \bibinfo{year}{2024}\natexlab{}.
\newblock \bibinfo{booktitle}{\emph{Applications of Python (Explained with Examples)}}.
\newblock
\urldef\tempurl%
\url{https://www.simplilearn.com/what-is-python-used-for-article}
\showURL{%
\tempurl}


\bibitem[Snyk(2024)]%
        {snyk_code}
\bibfield{author}{\bibinfo{person}{Snyk}.} \bibinfo{year}{2024}\natexlab{}.
\newblock \bibinfo{booktitle}{\emph{Snyk Code: Secure Your Code as You Develop}}.
\newblock
\urldef\tempurl%
\url{https://snyk.io/product/snyk-code}
\showURL{%
\tempurl}


\bibitem[Sobania et~al\mbox{.}(2022)]%
        {sobania2022choose}
\bibfield{author}{\bibinfo{person}{Dominik Sobania}, \bibinfo{person}{Martin Briesch}, {and} \bibinfo{person}{Franz Rothlauf}.} \bibinfo{year}{2022}\natexlab{}.
\newblock \showarticletitle{Choose your programming copilot: a comparison of the program synthesis performance of github copilot and genetic programming}. In \bibinfo{booktitle}{\emph{Proceedings of the 24th Annual Conference on Genetic and Evolutionary Computation Conference (GECCO)}}. ACM, \bibinfo{pages}{1019--1027}.
\newblock
\urldef\tempurl%
\url{https://doi.org/10.1145/3512290.3528700}
\showDOI{\tempurl}


\bibitem[Sui et~al\mbox{.}(2020)]%
        {sui2020recall}
\bibfield{author}{\bibinfo{person}{Li Sui}, \bibinfo{person}{Jens Dietrich}, \bibinfo{person}{Amjed Tahir}, {and} \bibinfo{person}{George Fourtounis}.} \bibinfo{year}{2020}\natexlab{}.
\newblock \showarticletitle{On the recall of static call graph construction in practice}. In \bibinfo{booktitle}{\emph{Proceedings of the 42nd ACM/IEEE International Conference on Software Engineering (ICSE)}}. ACM, \bibinfo{pages}{1049--1060}.
\newblock
\urldef\tempurl%
\url{https://doi.org/10.1145/3377811.3380441}
\showDOI{\tempurl}


\bibitem[Tahir and MacDonell(2012)]%
        {tahir2012systematic}
\bibfield{author}{\bibinfo{person}{Amjed Tahir} {and} \bibinfo{person}{Stephen~G MacDonell}.} \bibinfo{year}{2012}\natexlab{}.
\newblock \showarticletitle{A systematic mapping study on dynamic metrics and software quality}. In \bibinfo{booktitle}{\emph{2012 28th IEEE International Conference on Software Maintenance (ICSM)}}. IEEE, \bibinfo{pages}{326--335}.
\newblock
\urldef\tempurl%
\url{https://doi.org/10.1109/ICSM.2012.6405289}
\showDOI{\tempurl}


\bibitem[{The MITRE Corporation}(2023)]%
        {cwe-software-development}
\bibfield{author}{\bibinfo{person}{{The MITRE Corporation}}.} \bibinfo{year}{2023}\natexlab{}.
\newblock \bibinfo{booktitle}{\emph{CWE VIEW: Software Development}}.
\newblock
\urldef\tempurl%
\url{https://cwe.mitre.org/data/definitions/699.html}
\showURL{%
\tempurl}


\bibitem[T{\'o}masd{\'o}ttir et~al\mbox{.}(2018)]%
        {tomasdottir2018adoption}
\bibfield{author}{\bibinfo{person}{Krist{\'\i}n~Fj{\'o}la T{\'o}masd{\'o}ttir}, \bibinfo{person}{Mauricio Aniche}, {and} \bibinfo{person}{Arie Van~Deursen}.} \bibinfo{year}{2018}\natexlab{}.
\newblock \showarticletitle{The adoption of javascript linters in practice: A case study on eslint}.
\newblock \bibinfo{journal}{\emph{IEEE Transactions on Software Engineering}} \bibinfo{volume}{46}, \bibinfo{number}{8} (\bibinfo{year}{2018}), \bibinfo{pages}{863--891}.
\newblock
\urldef\tempurl%
\url{https://doi.org/10.1109/TSE.2018.2871058}
\showDOI{\tempurl}


\bibitem[Tony et~al\mbox{.}(2023)]%
        {tony2023llmseceval}
\bibfield{author}{\bibinfo{person}{Catherine Tony}, \bibinfo{person}{Markus Mutas}, \bibinfo{person}{Nicol{\'a}s E~D{\'\i}az Ferreyra}, {and} \bibinfo{person}{Riccardo Scandariato}.} \bibinfo{year}{2023}\natexlab{}.
\newblock \showarticletitle{Llmseceval: A dataset of natural language prompts for security evaluations}. In \bibinfo{booktitle}{\emph{2023 IEEE/ACM 20th International Conference on Mining Software Repositories (MSR)}}. IEEE, \bibinfo{pages}{588--592}.
\newblock
\urldef\tempurl%
\url{https://doi.org/10.1109/MSR59073.2023.00084}
\showDOI{\tempurl}


\bibitem[Vaithilingam et~al\mbox{.}(2022)]%
        {vaithilingam2022expectation}
\bibfield{author}{\bibinfo{person}{Priyan Vaithilingam}, \bibinfo{person}{Tianyi Zhang}, {and} \bibinfo{person}{Elena~L Glassman}.} \bibinfo{year}{2022}\natexlab{}.
\newblock \showarticletitle{Expectation vs. experience: Evaluating the usability of code generation tools powered by large language models}. In \bibinfo{booktitle}{\emph{Proceedings of the 40th ACM Conference on Human Factors in Computing Systems (CHI)}}. ACM, \bibinfo{pages}{1--7}.
\newblock
\urldef\tempurl%
\url{https://doi.org/10.1145/3491101.3519665}
\showDOI{\tempurl}


\bibitem[Waldinger and Lee(1969)]%
        {waldinger1969prow}
\bibfield{author}{\bibinfo{person}{Richard~J Waldinger} {and} \bibinfo{person}{Richard~CT Lee}.} \bibinfo{year}{1969}\natexlab{}.
\newblock \showarticletitle{PROW: A step toward automatic program writing}. In \bibinfo{booktitle}{\emph{Proceedings of the 1st International Joint Conference on Artificial Intelligence (IJCAI)}}. ACM, \bibinfo{pages}{241--252}.
\newblock


\bibitem[White et~al\mbox{.}(2023)]%
        {white2023prompt}
\bibfield{author}{\bibinfo{person}{Jules White}, \bibinfo{person}{Quchen Fu}, \bibinfo{person}{Sam Hays}, \bibinfo{person}{Michael Sandborn}, \bibinfo{person}{Carlos Olea}, \bibinfo{person}{Henry Gilbert}, \bibinfo{person}{Ashraf Elnashar}, \bibinfo{person}{Jesse Spencer-Smith}, {and} \bibinfo{person}{Douglas~C Schmidt}.} \bibinfo{year}{2023}\natexlab{}.
\newblock \showarticletitle{A prompt pattern catalog to enhance prompt engineering with chatgpt}.
\newblock \bibinfo{journal}{\emph{arXiv preprint arXiv:2302.11382}} (\bibinfo{year}{2023}).
\newblock
\urldef\tempurl%
\url{https://doi.org/10.48550/arXiv.2302.11382}
\showDOI{\tempurl}


\bibitem[Wong et~al\mbox{.}(2022)]%
        {wong2022exploring}
\bibfield{author}{\bibinfo{person}{Dakota Wong}, \bibinfo{person}{Austin Kothig}, {and} \bibinfo{person}{Patrick Lam}.} \bibinfo{year}{2022}\natexlab{}.
\newblock \showarticletitle{Exploring the Verifiability of Code Generated by GitHub Copilot}.
\newblock \bibinfo{journal}{\emph{arXiv preprint arXiv:2209.01766}} (\bibinfo{year}{2022}).
\newblock
\urldef\tempurl%
\url{https://doi.org/10.48550/arXiv.2209.01766}
\showDOI{\tempurl}


\bibitem[Yeti{\c{s}}tiren et~al\mbox{.}(2023)]%
        {yeticstiren2023evaluating}
\bibfield{author}{\bibinfo{person}{Burak Yeti{\c{s}}tiren}, \bibinfo{person}{I{\c{s}}{\i}k {\"O}zsoy}, \bibinfo{person}{Miray Ayerdem}, {and} \bibinfo{person}{Eray T{\"u}z{\"u}n}.} \bibinfo{year}{2023}\natexlab{}.
\newblock \showarticletitle{Evaluating the Code Quality of AI-Assisted Code Generation Tools: An Empirical Study on GitHub Copilot, Amazon CodeWhisperer, and ChatGPT}.
\newblock \bibinfo{journal}{\emph{arXiv preprint arXiv:2304.10778}} (\bibinfo{year}{2023}).
\newblock
\urldef\tempurl%
\url{https://doi.org/10.48550/arXiv.2304.10778}
\showDOI{\tempurl}


\bibitem[Yetistiren et~al\mbox{.}(2022)]%
        {yetistiren2022assessing}
\bibfield{author}{\bibinfo{person}{Burak Yetistiren}, \bibinfo{person}{Isik Ozsoy}, {and} \bibinfo{person}{Eray Tuzun}.} \bibinfo{year}{2022}\natexlab{}.
\newblock \showarticletitle{Assessing the quality of GitHub copilot’s code generation}. In \bibinfo{booktitle}{\emph{Proceedings of the 18th International Conference on Predictive Models and Data Analytics in Software Engineering (PROMISE)}}. ACM, \bibinfo{pages}{62--71}.
\newblock
\urldef\tempurl%
\url{https://doi.org/10.1145/3558489.355907}
\showDOI{\tempurl}


\bibitem[Yu et~al\mbox{.}(2024)]%
        {yu2024large}
\bibfield{author}{\bibinfo{person}{Xiao Yu}, \bibinfo{person}{Lei Liu}, \bibinfo{person}{Xing Hu}, \bibinfo{person}{Jacky~Wai Keung}, \bibinfo{person}{Jin Liu}, {and} \bibinfo{person}{Xin Xia}.} \bibinfo{year}{2024}\natexlab{}.
\newblock \showarticletitle{Where Are Large Language Models for Code Generation on GitHub?}
\newblock \bibinfo{journal}{\emph{arXiv preprint arXiv:2406.19544}} (\bibinfo{year}{2024}).
\newblock
\urldef\tempurl%
\url{https://doi.org/10.48550/arXiv.2406.19544}
\showDOI{\tempurl}


\bibitem[Zhang et~al\mbox{.}(2023)]%
        {zhang2023demystifying}
\bibfield{author}{\bibinfo{person}{Beiqi Zhang}, \bibinfo{person}{Peng Liang}, \bibinfo{person}{Xiyu Zhou}, \bibinfo{person}{Aakash Ahmad}, {and} \bibinfo{person}{Muhammad Waseem}.} \bibinfo{year}{2023}\natexlab{}.
\newblock \showarticletitle{Demystifying Practices, Challenges and Expected Features of Using GitHub Copilot}.
\newblock \bibinfo{journal}{\emph{International Journal of Software Engineering and Knowledge Engineering}} \bibinfo{volume}{33}, \bibinfo{number}{11\&12} (\bibinfo{year}{2023}), \bibinfo{pages}{1653--1672}.
\newblock
\urldef\tempurl%
\url{https://doi.org/10.1142/S0218194023410048}
\showDOI{\tempurl}


\end{thebibliography}

\end{document}